\documentclass[12pt]{article}
\usepackage{amsmath,amssymb}
\usepackage{bm}
\usepackage{color}
\usepackage{graphicx}
\usepackage{cite}
\usepackage{mathtools}
\usepackage{breqn}
\usepackage{braket, multirow, subfigure, multicol}
\usepackage{here}

\newcommand{\maru}[1]{\raise0.2ex\hbox{\textcircled{\scriptsize{#1}}}}

\begin{document}
\title{
\begin{flushright}
\ \\*[-80pt]
\begin{minipage}{0.2\linewidth}
\normalsize
%arXiv:YYMM.NNNN \\
EPHOU-22-004\\*[50pt]
\end{minipage}
\end{flushright}
% Title
{\Large \bf
Classifications of magnetized $T^4$\\ and $T^4/{Z}_2$ orbifold models
\\*[20pt]}}
% /Title

\author{
~Shota Kikuchi,
%\footnote{B's mail}
~Tatsuo Kobayashi, 
%\footnote{C's mail}
~Kaito Nasu, and 
%\footnote{D's mail}
~Hikaru Uchida
%\footnote{E's mail}
\\*[20pt]
\centerline{
\begin{minipage}{\linewidth}
\begin{center}
{\it \normalsize
Department of Physics, Hokkaido University, Sapporo 060-0810, Japan} \\*[5pt]
\end{center}
\end{minipage}}
\\*[50pt]}

\date{
\centerline{\small \bf Abstract}
\begin{minipage}{0.9\linewidth}
\medskip
\medskip
\small
We study constructions and classifications of three-generation models based on magnetized $T^4$ and $T^4/{Z}_2$ orbifold as candidates of the compact space. We focus on chiral fermion zero-mode wave functions in the extra dimensions. Freedoms of constant gauge fields, called Scherk-Schwarz phases are taken into account. 
%Infinite number of three-generation models are yielded, corresponding to the ways in which the magnetic flux can be turned on. 
We classify three-generation models in a systematic manner, clarifying the relationship between different models. The Higgs sector is also studied by analyzing possible assignments of the magnetic flux and Scherk-Schwarz phases, etc. to left- and right-handed fermions.   
\end{minipage}
}

\begin{titlepage}
\maketitle
\thispagestyle{empty}
\end{titlepage}

\newpage

% ------------------------------------------------------ %
% ------------------------------------------------------ %
% ------------------------------------------------------ %
% ------------------------------------------------------ %
%\setlength{\tabcolsep}{0.8mm} 
\section{Introduction}
\label{Intro}

The Standard Model (SM) of particle physics leave some fundamental questions of the universe unanswered. We wonder why three generations of quarks and leptons appear. The observed parameters such as masses and mixing angles among the generations of particles show some remarkable features without clear reasons. For instance, we can see large mass hierarchies in both lepton and quark sectors. It is also mysterious why the degree of mixings is so different between the two sectors. The SM cannot explain these behaviors. Furthermore, the origin of the gauge group structure $SU(3) \times SU(2) \times U(1)$ is unclear. The SM is a chiral theory that left- and right- handed fermions have different quantum numbers and representations.  

In order to answer these questions, one should look for a high-energy underlying theory. Superstring theory is a promising candidate. This theory predicts 10 dimensional spacetime, so the extra 6 dimensional space needs to be compactified. 
The gauge group structure originates from D-brane configurations in the compact space.
(See for reviews Refs.~\cite{Blumenhagen:2006ci,Ibanez:2012zz}.) 
In addition, the geometry of the compact space determines the flavor structures and the patterns of the above-mentioned parameters in the four dimensional (4D) theory.
%In addition, ideas to explain the gauge group structure were proposed by considering specific orientations and intersections of D6 branes based on the type II-A picture.

For quantitative evaluations of the relationship between the SM and superstring theory, one should first consider a relatively simple setup allowing analytical computations. 
An explicit way to compute Yukawa couplings was proposed based on the supersymmetric Yang-Mills theory (SYM) 
in higher dimensions known as a low-energy effective field theory of superstring theory. Yukawa couplings can be obtained by overlap integrals of zero-mode wave functions 
corresponding to the matter modes and the Higgs boson over the compact space. 
The torus compactifiction with magnetic fluxes is rather simple, but quite interesting ~\cite{Bachas:1995ik,Blumenhagen:2000wh,Angelantonj:2000hi,Blumenhagen:2000ea,Cremades:2004wa}.
That can lead to a 4D chiral theory, and one can realize realistic gauge symmetry breaking depending on configurations of 
the background magnetic fluxes.
Moreover, the strength of the magnetic flux determines the generation number. 
For the factorizable torus $T^2 \times T^2 \times T^2$ with magnetic fluxes, one can obtain explicit forms of the wave functions in terms of the Jacobi theta-function, and then one can compute Yukawa couplings and higher order 
couplings analytically \cite{Cremades:2004wa,Abe:2009dr}. 
The flavor structure such as mass hierarchies and mixing anlges is originated from wave functions on one of $T^2$'s, 
because the generation number, 3, is a prime number, although wave functions on other $T^2\times T^2$ 
contribute to only the overall factor of Yukawa matrices.

These studies were extended to the $T^2/Z_N$ orbifolds with magnetic fluxes \cite{Abe:2008fi,Abe:2013bca,Abe:2014noa,Kobayashi:2017dyu}.
We have various possibilities to construct three-generation models on the $T^2/Z_N$ orbifolds, 
and they have a rich structure in the flavor sector.
Three-generation models were classified \cite{Abe:2008sx,Abe:2015yva}.
Yukawa couplings were studied, and realistic masses and mixing angles were realized \cite{Abe:2012fj,Abe:2014vza,Fujimoto:2016zjs,Kobayashi:2016qag,Kikuchi:2021yog}.
Their flavor structure is controlled by the flavor symmetry \cite{Abe:2009vi} including the modular flavor symmetry 
\cite{Kobayashi:2018rad,Kobayashi:2018bff,Kariyazono:2019ehj,Ohki:2020bpo,Kikuchi:2020frp,Kikuchi:2020nxn,
Kikuchi:2021ogn,Almumin:2021fbk,Kikuchi:2022bkn}.

%Through the  magnetized $T^2 \times T^2 \times T^2$ compactification, one can indeed obtain explicit form of the wavefunctions in terms of Jacobi theta function \cite{Cremades:2004wa}. Thus, numerous studies were carried out under this setup. Some variations involving orbifolds $T^2/{Z}_N,\ (N=2,3,4,6)$ were also studied \cite{Abe:2013bca}. It should be emphasized that depending on the orientations of the background magnetic fluxes, one can realize some realistic gauge symmetry breakings as well as the chiral structure in the 4D-theory. Moreover, the strength of the magnetic flux determines the generation number. Thus, the analysis of SYM on magnetized torus is a highly motivated study. 

However, reminding the fact that the extra space is 6 dimensional, it is clearly insufficient to study the factorized case $T^2 \times T^2 \times T^2$, that is when the corresponding complex structure parameter $\Omega$ is diagonal. For example, we should extend the analysis to non-factorizable tori $T^4$ or $T^6$ whose complex structures are not necessarily diagonal. In fact, zero-mode wave functions for those generalized cases have already been proposed in terms of the Riemann theta-function \cite{Cremades:2004wa}. Moreover, corresponding theta identities were revealed which are crucial for the analytical evaluation of Yukawa couplings \cite{Antoniadis:2009bg}. 

There is a distinct difference between $T^{2g}, (g=2,3)$ and $T^2$ setups. In the latter case, the number of models yielding the three generations is limited very much. On the other hand, we have a number of three-generation models in the former case.
In string theory, the tadpole conditions constrain a full theory.
Here, we concentrate on the left-handed and right-handed chiral fermion sector,
 but do not study a full theory imposing the stringy consistency.
At any rate, we have many ways to choose  
%This is because we have infinitely many ways of choosing
 the orientations of the background magnetic flux $\bf{N}$. 
That is the infinite number in effective field theory if we do not impose the stringy consistency.
In order to handle this situation, an idea to classify $T^4$ models was suggested \cite{Abe:2014nla}. Models were classified into types, 
depending the flavor structure, which corresponds to alignment patterns of quasi-localized positions of  zero-mode wave functions. The proposed method is effective when the determinant of the magnetic flux ${\rm det} {\bf{N}}$ is either small or takes prime numbers.
It was hypothesized that the number of types is equal to the sum of all divisors of ${\rm det} {\bf{N}}$ without a rigorous proof. 

In this paper, we propose a different method of classifications of three-generation models defined on $T^4$. 
Our method is equivalent to the previous one \cite{Abe:2014nla}, which corresponds to the classification of 
the alignment pattern of quasi-localizing points of zero-mode wave functions.
We extend it to $T^4/Z_2$ orbifold models as well.
Our classification is also useful to study $T^4/Z_2$ models, 
because the number of zero-modes on the orbifold depends on the alignment pattern 
of quasi-localizing points of zero-mode wave functions on $T^4$.
That is, different Types lead to different numbers of zero-mode wave functions, in general.
Using our method, we can study all the possibilities of the background magnetic flux $\bf{N}$ to lead 
to three generations on $T^4/Z_2$.

 This new approach offers group theoretical point of views allowing simple understanding even when the size of magnetic flux is arbitrary. It also shows consistencies with the results in the preceding studies, and succeeds in proving the hypothesis. 

This paper is organized as follows. In section 2, we give a brief review on magnetized $T^4$ models, presenting the zero-mode wave functions. After that, we take into account Scherk-Schwarz phases and consider $T^4/Z_2$ orbifold models. In section 3, we present and discuss the classification of three-generation models on $T^4$, making a reference to supersymmetry (SUSY) condition as well. In section 4, we study models on $T^4/Z_2$, and clarify how three-generation models can be constructed. We also give a systematic classification of $T^4/Z_2$ models. In section 5, the Higgs sector is studied based on $T^4/Z_2$ models. 
In section 6, we explain how our way of classification is reflected in the flavor structure of the Yukawa coupling constants in magnetized $T^4$ models. Section 7 is our conclusion. In Appendix A, we give a derivation of the theta identity, and show Yukawa couplings in magnetized $T^4$ models including  Scherk-Schwarz phases. In Appendix B, we give proofs of some facts we used in Section 6.

% ------------------------------------------------------ %
% ------------------------------------------------------ %
% ------------------------------------------------------ %
% ------------------------------------------------------ %

\section{Compactification with magnetic fluxes}
\label{sec:Revew}

Here we review zero-mode wave functions on magnetized $T^4$ and $T^4/Z_2$ \cite{Cremades:2004wa}, \cite{Antoniadis:2009bg}. In order to simplify the discussion we focus on the case when the gauge symmetry is just $U(1)$.    

\subsection{Magnetic flux on $T^4$ }
We begin with writing down the magnetic field and the corresponding gauge potential on $T^4$.
$T^4$ can be defined as follows.
We denote the four-dimensional coordinates by $\vec{z} = (z^1, z^2) \in \mathbb{C}^2$.
Let $\Lambda$ be a lattice spanned by 4 vectors 
\begin{align}
\begin{aligned}
\label{eq: T^4_def}
\vec{e}_1 &= (1,0),\quad \Omega \vec{e}_1 &= (\Omega_{11}, \Omega_{12}), \\
\vec{e}_2 &= (0,1), \quad \Omega \vec{e}_2 &= (\Omega_{21}, \Omega_{22}),
\end{aligned}
\end{align}
where $\Omega$ is a $2 \times 2$ complex matrix and satisfies ${\rm Im}\Omega >0$, namely its eigenvalues are greater than zero. 
Then we define $T^4 \simeq \mathbb{C}^2/\Lambda$. This means we have the following identifications on $\mathbb{C}^2$,
\begin{equation}
\vec{z} \sim \vec{z} + \vec{e}_i \sim \vec{z} + \Omega \vec{e}_i .
\end{equation}
The background magnetic flux on $T^4$ is written by
\begin{equation}
F = \frac{1}{2} p_{x^ix^j} dx^i \wedge dx^j + \frac{1}{2} p_{y^i y^j} dy^i \wedge dy^j + p_{x^i y^j} dx^i \wedge dy^j.
\end{equation}
In terms of the complex coordinates,
\begin{equation}
z^i = x^i + \Omega^i_{\ j} y^j,\quad \bar{z}^i = x^i + \bar{\Omega}^i_{\ j} y^j,
\end{equation}
the flux is given by
\begin{equation}
F = \frac{1}{2} F_{z^i z^j} dz^i \wedge dz^j + \frac{1}{2} F_{\bar{z}^i \bar{z}^j} d\bar{z}^i \wedge d\bar{z}^j + F_{z^i \bar{z}^j} (i dz^i \wedge d\bar{z}^j) ,
\end{equation}
where
\begin{align}
F_{z^i z^j} &= {(\bar{\Omega} - \Omega)^{-1}}^T (\bar{\Omega}^T p_{xx} \bar{\Omega} + p_{yy} + p_{xy}^T \bar{\Omega} - \bar{\Omega}^T p_{xy}) {(\bar{\Omega} - \Omega)^{-1}}, \notag\\
F_{\bar{z}^i \bar{z}^j} &= {(\bar{\Omega} - \Omega)^{-1}}^T ({\Omega}^T p_{xx} {\Omega} + p_{yy} + p_{xy}^T {\Omega} - {\Omega}^T p_{xy}) {(\bar{\Omega} - \Omega)^{-1}}, \\
F_{z^i \bar{z}^j} &= i {(\bar{\Omega} - \Omega)^{-1}}^T \left( \bar{\Omega}^T p_{xx} \Omega + p_{yy}+ p_{xy}^T \Omega - \bar{\Omega}^T p_{xy}  \right) {(\bar{\Omega} - \Omega)^{-1}}. \notag
\end{align}
From the requirement that the supersymmetry survives, the flux needs to be a $(1,1)$-form. This can be satisfied by demanding the following condition
\begin{equation}
{\Omega}^T p_{xx} {\Omega} + p_{yy} + p_{xy}^T {\Omega} - {\Omega}^T p_{xy} = 0.
\end{equation}
As a result, $F$ can be written as
\begin{equation}
F = i (p_{xx} \Omega - p_{xy}) (\bar{\Omega} - \Omega)^{-1} (i dz^i \wedge d\bar{z}^j)  = -i {(\bar{\Omega}-\Omega)^{-1}}^T(\bar{\Omega}^T p_{xx} + p_{xy}^T) (i dz^i \wedge d\bar{z}^j) . 
\end{equation}
From this, we see the hermiticity of the flux: $F_{z^i \bar{z}^j}=F^{\dagger}_{z^i \bar{z}^j}$. 
The expression is simplified further by assuming $p_{xx}=p_{yy}=0$,
\begin{equation}
F_{z^i \bar{z}^j} = i [p_{xy} (\Omega - \bar{\Omega})^{-1}]_{ij} .
\end{equation}
The SUSY condition is also simplified
\begin{equation}
(p_{xy}^T \Omega)^T = p_{xy}^T \Omega.
\end{equation}
The flux must be quantized due to the consistency with the boundary conditions we take. Thus, we can write $p_{xy}$ in terms of a $2 \times 2$ real integer matrix $\bf{N}$, referred as the intersection matrix,
\begin{equation}
p_{xy} = 2 \pi {\bf{N}}^T .
\end{equation}
This is referred to as Dirac's quantization condition. 
Then we obtain
\begin{equation}
F = \pi \left[{\bf{N}}^T ({\rm Im}\Omega)^{-1} \right]_{ij} (i dz^i \wedge d\bar{z}^j),
\end{equation}
with the SUSY condition rewritten as
\begin{equation}
({\bf N} \Omega)^T = {\bf{N}} \Omega.
\end{equation}
The corresponding gauge potential is given by
\begin{align}
\begin{aligned}
A(\vec{z}, \vec{\bar{z}}) &= \pi  {\rm Im} \left\{ [{\bf{N}} (\vec{\bar{z}}  + \vec{\bar{\zeta}})] ({\rm Im} \Omega)^{-1} d \vec{z}  \right\}  \\
&= - \frac{\pi i}{2} \left\{ [{\bf{N}} (\vec{\bar{z}}  + \vec{\bar{\zeta}})] ({\rm Im} \Omega)^{-1} \right\}_i dz^i + \frac{\pi i}{2} \left\{ [{\bf{N}} (\vec{{z}}  + \vec{{\zeta}})] ({\rm Im} \Omega)^{-1} \right\}_i d\bar{z}^i \\
& =: A_{z^i} dz^i + A_{\bar{z}^i} d\bar{z}^i,
\end{aligned}
\end{align}
where $\vec{\zeta}$ is a complex constant 2-component vector known as the Wilson lines. The boundary conditions are
\begin{align}
\begin{aligned}
\label{eq: A_boundary}
A(\vec{z}+\vec{e}_k) &= A(\vec{z}) + d \chi_{\vec{e}_k}(\vec{z}), \\
A(\vec{z} + \Omega \vec{e}_k) &= A(\vec{z}) + d\chi_{\Omega \vec{e}_k} (\vec{z}), 
\end{aligned}
\end{align}
where 
\begin{align}
\begin{aligned}
\chi_{\vec{e}_k} (\vec{z}) := \pi [{\bf{N}}^T ({\rm Im}\Omega)^{-1} {\rm Im} (\vec{z} + \vec{\zeta})]_k,&\quad \chi_{\Omega \vec{e}_k} (\vec{z}) := \pi {\rm Im}[{\bf{N}} \bar{\Omega} ({\rm Im}\Omega)^{-1} (\vec{z} + \vec{\zeta})]_k.
\end{aligned}
\end{align}
Since the Wilson lines are equivalent to the Scherk-Schwarz (SS) phases by the $U(1)$ gauge transformation \cite{Abe:2013bca},
 we take $\vec{\zeta}=0$ from now on.\footnote{
The Wilson lines correspond to the D-brane position moduli and the open string moduli in the T-dual of magnetized D-brane models.}

\subsection{Zero-mode wave function without SS phases}
\subsubsection{$T^4$}
We first consider the case under the positive definite condition
\begin{equation}
{\bf{N}} \cdot {\rm Im}\Omega > 0.
\end{equation}
This means we have ${\rm det}{\bf{N}} > 0$ and ${\rm Tr} ({\bf{N}} \cdot {\rm Im}\Omega)>0$. Then the following wave functions on $T^4$:
\begin{equation}
\label{eq: zero-mode}
\psi^{\vec{j}} (\vec{z}, \Omega) = \mathcal{N} e^{\pi i [{\bf{N}}\vec{z}] \cdot ({\rm Im}\Omega)^{-1} \cdot {\rm Im}\vec{z} } \cdot \vartheta
 \begin{bmatrix}
 \vec{j}  \\
 0
 \end{bmatrix}({\bf{N}} \vec{z}, {\bf{N}}\Omega),\quad \vec{j} \cdot {\bf{N}} \in \mathbb{Z}^2,
\end{equation}
are well-defined, where 
\begin{equation}
\vartheta 
 \begin{bmatrix}
 \vec{a} \\
 \vec{b}
 \end{bmatrix} (\vec{\nu}, \tilde{\Omega}) = 
 \sum_{\vec{l} \in \mathbb{Z}^2} e^{\pi i (\vec{l} + \vec{a}) \cdot \tilde{\Omega} \cdot (\vec{l} + \vec{a})} e^{2 \pi i (\vec{l} + \vec{a}) \cdot (\vec{\nu} + \vec{b} )},\quad \vec{a}, \vec{b} \in \mathbb{R}^2,\ \tilde{\Omega}^T = \tilde{\Omega},\ {\rm Im} {\tilde{\Omega}} > 0,
\end{equation}
is known as the Riemann theta-function. Note that $\vec{j}$ is defined up to integer because $\psi^{\vec{j}} = \psi^{\vec{j}+ \vec{e}_k}$. $\mathcal{N}$ is just a normalization constant. 
$\psi^{\vec{j}}$ are the zero-mode wave functions of the Dirac equation,
\begin{align}
D_i \psi^{\vec{j}, {\bf{N}}}(\vec{z}, \Omega) = 0, \quad (i=1, 2) , \notag \\
D_i = \bar{\partial}_i + \frac{\pi}{2} \left( [{\bf{N}} \vec{z} ] \cdot ({\rm Im}\Omega)^{-1} \right)_i .
\end{align}
The wave functions $\psi^{\vec{j}}$ satisfy the following boundary conditions:
\begin{align}
\begin{aligned}
\label{eq: boundary1}
\psi^{\vec{j}} (\vec{z} + \vec{e}_k, \Omega) &= e^{i \chi_{\vec{e}_k} (\vec{z})} \cdot \psi^{\vec{j}}(\vec{z}, \Omega), \\
\psi^{\vec{j}} (\vec{z} + \Omega \vec{e}_k, \Omega) &=  e^{i \chi_{\Omega \vec{e}_k} (\vec{z})} \cdot \psi^{\vec{j}}(\vec{z}, \Omega).
\end{aligned}
\end{align}
We can clearly see consistencies with the boundary conditions of the gauge potential shown by Eq.($\ref{eq: A_boundary}$). Finally, it is important to note that the degeneracy of the zero-modes is given by the determinant of ${\bf{N}}$. 

When ${\bf{N}} \cdot {\rm Im}\Omega < 0$, we have ${\rm det}{\bf{N}}>0$ and ${\rm Tr}({\bf{N}}\cdot {{\rm Im} \Omega}) < 0$. Then, wave functions in Eq.$(\ref{eq: zero-mode})$ are not defined. In this case, we need to take 
\begin{equation}
    \psi^{\vec{j}} (\vec{\bar{z}}, \bar{\Omega}) = \mathcal{N} e^{\pi i [{\bf{N}}\vec{\bar{z}}] \cdot ({\rm Im}\bar{\Omega})^{-1} \cdot {\rm Im}\vec{\bar{z}} } \cdot \vartheta
 \begin{bmatrix}
 \vec{j}  \\
 0
 \end{bmatrix}({\bf{N}} \vec{\bar{z}}, {\bf{N}}\bar{\Omega}).
\end{equation}
When ${\rm det}{\bf{N}} < 0$, we need to replace $\Omega$ by the effective complex structure $\Omega_{\rm eff.} = \hat{\Omega} \Omega$ as stated in Ref. \cite{Antoniadis:2009bg}, 
\begin{equation}
    \hat{\Omega} := \frac{1}{1+q^2} 
    \begin{pmatrix}
    1 - q^2 & - 2 q \\
    - 2 q & q^2 -1
    \end{pmatrix} ,
\end{equation}
where $q$ is given by the following relation
\begin{equation}
    {\bf{N}}= 
     \begin{pmatrix}
     n_{11} & n_{12} \\
     n_{21} & n_{22} 
     \end{pmatrix}
     = n_{11} 
     \begin{pmatrix}
     1 & -q \\
     -q & q^2
     \end{pmatrix} 
     + n_{22}
     \begin{pmatrix}
     q^2 & q \\
     q & 1
     \end{pmatrix} .
\end{equation}

In order to make the following discussions simple, let us concentrate on the case when wave functions are given by Eq.($\ref{eq: zero-mode}$), namely positive chirality modes with the positive definite condition ${\bf{N}} \cdot {\rm Im} \Omega >0$. 

\subsubsection{$T^4/Z_2$}
We define $T^4/Z_2$ orbifold by imposing a further identification to $T^4$. In addition to Eq.(\ref{eq: T^4_def}), we demand
\begin{equation}
\vec{z} \sim - \vec{z}. 
\end{equation}
Then the zero-mode wave functions must be either $Z_2$ even $\psi_{T^4/Z_2, +}$ or odd $\psi_{T^4/Z_2, -}$,
\begin{equation}
\psi_{T^4/Z_2, \pm} (- \vec{z}, \Omega) = \pm \psi_{T^4/Z_2, \pm} ( \vec{z}, \Omega).
\end{equation}
We can construct them by the linear combinations of zero-mode wave functions on $T^4$,
\begin{align}
\psi_{T^4/Z_2, +}^{\vec{j}} (\vec{z}, \Omega) &\propto \psi^{\vec{j}} (\vec{z}, \Omega) + \psi^{\vec{j}} (-\vec{z}, \Omega), \notag \\
\psi_{T^4/Z_2, -}^{\vec{j}} (\vec{z}, \Omega) &\propto \psi^{\vec{j}} (\vec{z}, \Omega) - \psi^{\vec{j}} (-\vec{z}, \Omega).
\end{align}
The following relation is useful,
\begin{equation}
\psi^{\vec{j}} (-\vec{z}, \Omega) = \psi^{\vec{e} -\vec{j}} (\vec{z}, \Omega) = \psi^{-\vec{j}} (\vec{z}, \Omega),
\end{equation}
where $\vec{e}=\vec{e}_1+\vec{e}_2$. In particular, when $\vec{j} = (0,0), (0, \frac{1}{2}), (\frac{1}{2}, 0), (\frac{1}{2}, \frac{1}{2})$, 
wave functions  satisfy $\psi^{\vec{j}} (- \vec{z}, \Omega) = \psi^{\vec{j}} (\vec{z}, \Omega)$, and we refer to them as invariant modes.

\subsection{Zero-mode wave function with SS phases}
\subsubsection{$T^4$}
Here we write down zero-mode wave functions including Scherk-Schwarz phases on $T^4$. Boundary conditions Eq.($\ref{eq: boundary1}$) are generalized, and we demand
\begin{align}
\begin{aligned}
\label{eq: b_1}
\psi^{(\vec{j} + \vec{\alpha} \bm{N}^{-1}, \vec{\beta} )}(\vec{z} + \vec{e}_k, \Omega) &= e^{2 \pi i \alpha_k} \cdot  e^{i \chi_{\vec{e}_k} (\vec{z})} \cdot \psi^{( \vec{j} + \vec{\alpha} \bm{N}^{-1}, \vec{\beta} )}(\vec{z}, \Omega), \\
\psi^{( \vec{j} + \vec{\alpha} \bm{N}^{-1}, \vec{\beta} )}(\vec{z} + \Omega \vec{e}_k, \Omega) &= e^{2 \pi i \beta_k} \cdot  e^{i \chi_{\Omega \vec{e}_k} (\vec{z})} \cdot \psi^{( \vec{j} + \vec{\alpha} \bm{N}^{-1}, \vec{\beta} )}(\vec{z}, \Omega),
\end{aligned}
\end{align}
where $\vec{\alpha}$ and $\vec{\beta}$ are two dimensional real vectors and can take arbitrary values. We call them Scherk-Schwarz phases. The zero-mode wave functions satisfying the above boundary conditions are given by

\begin{equation}
\psi^{(\vec{j}+\vec{\alpha} {\bf{N}}^{-1}, \vec{\beta})} (\vec{z}, \Omega) = \mathcal{N} e^{\pi i [{\bf{N}}\vec{z}] \cdot ({\rm Im}\Omega)^{-1} \cdot {\rm Im}\vec{z} } \cdot \vartheta
 \begin{bmatrix}
 \vec{j} + \vec{\alpha} {\bf N}^{-1} \\
 - \vec{\beta}
 \end{bmatrix}({\bf{N}} \vec{z}, {\bf{N}}\Omega),
\end{equation}
where $\vec{j} \cdot {\bf{N}} \in \mathbb{Z}^2$.

\subsubsection{$T^4/Z_2$}
Here we study zero-mode wave functions on $T^4/{Z}_2$,
\begin{align}
\label{eq: Z_2_zero}
\psi_{T^4/Z_2, +}^{(\vec{j}+\vec{\alpha} {\bf{N}}^{-1}, \vec{\beta})} (\vec{z}, \Omega) &\propto \psi^{(\vec{j}+\vec{\alpha} {\bf{N}}^{-1}, \vec{\beta})}(\vec{z}, \Omega) + \psi^{(\vec{j}+\vec{\alpha} {\bf{N}}^{-1}, \vec{\beta})} (-\vec{z}, \Omega), \\
\label{eq: Z_2_zero_odd}
\psi_{T^4/Z_2, -}^{(\vec{j}+\vec{\alpha} {\bf{N}}^{-1}, \vec{\beta})} (\vec{z}, \Omega) &\propto \psi^{(\vec{j}+\vec{\alpha} {\bf{N}}^{-1}, \vec{\beta})} (\vec{z}, \Omega) - \psi^{(\vec{j}+\vec{\alpha} {\bf{N}}^{-1}, \vec{\beta})} (-\vec{z}, \Omega),
\end{align}
where $\psi_{T^4/Z_2, \pm}(\vec{z}, \Omega) $ must satisfy boundary conditions of $T^4$. Thus, $\psi^{(\vec{j} + \vec{\alpha} {\bf{N}}^{-1}, \vec{\beta})} (-\vec{z}, \Omega)$ must satisfy the conditions shown by Eq.($\ref{eq: b_1}$), i.e.,
\begin{align}
\begin{aligned}
\psi(-(\vec{z} + \vec{e}_k ), \Omega) = e^{- 2 \pi i \alpha_k} \cdot e^{i \chi_{\vec{e}_k} (\vec{z})} \cdot \psi^{(\vec{j} + \vec{\alpha} {\bf{N}}^{-1}, \vec{\beta})} (- \vec{z}, \Omega), \\
\psi(-(\vec{z} + \Omega \vec{e}_k ), \Omega) = e^{- 2 \pi i \beta_k} \cdot e^{i \chi_{\Omega \vec{e}_k} (\vec{z})} \cdot \psi^{(\vec{j} + \vec{\alpha} {\bf{N}}^{-1}, \vec{\beta})} (- \vec{z}, \Omega) .
\end{aligned}
\end{align}
As a result, we obtain $e^{-2\pi i \alpha_k} = e^{2 \pi i \alpha_k}$ and $e^{- 2 \pi i \beta_k} = e^{2 \pi i \beta_k}$. 
Thus, Scherk-Schwarz phases can take 16 different values:\footnote{The discrete Scherk-Schwarz phases are equivalent to 
discrete Wilson lines.
In addition, discretized Wilson lines correspond to the stabilization of the open string moduli 
on the orbifold in the T-dual of magnetized D-brane models \cite{Blumenhagen:2005tn}.}
\begin{align}
\begin{aligned}
\label{eq: SSphasesZ2}
\vec{\alpha} = \left( 0\ {\rm or}\ 1/2,\  0\ {\rm or}\ 1/2 \right),\quad {\rm mod}\ 1, \\
\vec{\beta} = \left( 0\ {\rm or}\ 1/2,\  0\ {\rm or}\ 1/2 \right),\quad {\rm mod}\ 1.
\end{aligned}
\end{align} 

The following relation is useful,
\begin{equation}
\label{eq: Z2_wave_flip}
\psi^{(\vec{j}+\vec{\alpha} {\bf{N}}^{-1}, \vec{\beta})} (-\vec{z}, \Omega) = \psi^{(\vec{e}-(\vec{j}+\vec{\alpha} {\bf{N}}^{-1}), -\vec{\beta})} (\vec{z}, \Omega) = e^{-4\pi i (\vec{j} + \vec{\alpha} {\bf{N}}^{-1})\cdot \vec{\beta}}  \psi^{(\vec{e}-(\vec{j}+\vec{\alpha} {\bf{N}}^{-1}), \vec{\beta})}  ,
\end{equation}
where $\vec{e} = \vec{e}_1 + \vec{e}_2$. The second equality in Eq. ($\ref{eq: Z2_wave_flip}$) holds only when Scherk-Schwarz phases are given by Eq.($\ref{eq: SSphasesZ2}$).

\section{Three-generation models on $T^4$}
\label{sec:}
In order to obtain three-generation models, we need 3 degenerated zero-modes. Thus, we demand 
\begin{equation}
{\rm det} {\bf{N}} = 3.
\end{equation}
Obviously we have infinite number of ${\bf{N}}$ satisfying this condition.
Stringy tadpole cancellation conditions may constrain the size of ${\bf{N}}$, although it would also depend on other sectors.
At any rate, we study the possibility of ${\bf{N}}$ from the viewpoint of effective field theory.
In order to get better understanding we define Types in the intersection matrix $\bf{N}$. Two matrices ${\bf{N}}_1$ and ${\bf{N}}_2$ are in the same Type if and only if they are related by
\begin{equation}
\label{eq: Type}
{\bf{N}}_1 \gamma = {\bf{N}}_2,\qquad \gamma \in SL(2,\mathbb{Z})=\Gamma.
\end{equation}
Here, $\Gamma = SL(2,\mathbb{Z})$ is defined as
\begin{equation}
SL(2, \mathbb{Z}) :=
  \left\{ 
   \begin{pmatrix}
    a & b \\
    c & d
   \end{pmatrix}: a, b, c, d \in \mathbb{Z}, ad -bc = 1
    \right\}.
\end{equation}
This group is generated by
\begin{equation}
\label{eq: SL2Z_generators}
S = 
\begin{pmatrix}
0 & -1 \\ 1 & 0
\end{pmatrix},\quad 
T =
\begin{pmatrix}
1 & 1 \\  0 & 1
\end{pmatrix}.
\end{equation}

We get 4 different Types whose representative matrices are given by
\begin{align}
\label{eq: D3type}
{\rm Type\ I}:  \begin{pmatrix} 3 & 0 \\ 0 & 1 \end{pmatrix},\ 
{\rm Type\ II}:  \begin{pmatrix} 3 & 1 \\ 0 & 1 \end{pmatrix},\ 
{\rm Type\ III}: \begin{pmatrix} 3 & 2 \\ 0 & 1 \end{pmatrix},\ 
{\rm Type\ IV}: \begin{pmatrix} 1 & 0 \\ 0 & 3 \end{pmatrix}. 
\end{align}

\subsection{Types}
Here we give a proof of Eq.$(\ref{eq: D3type})$, the classification of ${\bf{N}}$ into Types. For later convenience, let us consider the general case that $D := {\rm det}{\bf{N}}$ is not only 3, but can also take an arbitrary integer. We explicitly show the representative matrices for each Type and the number of Types is given by the sum of divisors of $D$.  

We write the matrix elements explicitly as 
\begin{equation}
{\bf{N}} = 
 \begin{pmatrix}
 n_1 & n_2 \\ n_3 & n_4
 \end{pmatrix}.
\end{equation}
Let us denote the greatest common divisor of $n_3$ and $n_4$ by $x(\geq 1)$,
\begin{equation}
x = {\rm gcd}(n_3, n_4) .
\end{equation}
Note that $x$ must be a divisor of $D$. We can relate $n_3$ and $n_4$ by
\begin{equation}
n_4 = q n_3 + r,
\end{equation}
where $q$ and $r$ are both integers. We can in general make $|r| \leq \frac{|n_3|}{2}$.
We can verify 
\begin{equation}
\label{eq: n3_r}
x =  {\rm gcd}(n_3, r) .
\end{equation}
Now we multiply an element of $\Gamma = SL(2,\mathbb{Z})$ to the right of ${\bf{N}}$
\begin{equation}
 \begin{pmatrix}
 \label{eq: process1}
 n_1 & n_2 \\ n_3 & n_4
 \end{pmatrix}
  \begin{pmatrix}
 1 & -q \\ 0 & 1
 \end{pmatrix}
 =
 \begin{pmatrix}
 n_1 & n_2 - q n_1 \\
 n_3 & r
 \end{pmatrix}.
\end{equation}
Then we multiply  $S \in \Gamma $ to the right
\begin{equation}
\label{eq: process2}
 \begin{pmatrix}
 n_1 & n_2 - q n_1 \\
 n_3 & r
 \end{pmatrix}
 \begin{pmatrix}
 0 & -1 \\
 1 & 0
 \end{pmatrix}
 =
  \begin{pmatrix}
 n_2 - q n_1 & -n_1 \\
 r & -n_3
 \end{pmatrix}.
\end{equation}
From Eq.$(\ref{eq: n3_r})$, we can write
\begin{equation}
-n_3 = q' r + r',
\end{equation}
where ${\rm gcd}(r, r')=x$ and $|r'| \leq \frac{|r|}{2}$. Thus, we can repeat the procedure shown by Eqs.$(\ref{eq: process1})$ and $(\ref{eq: process2})$. After some iterations and the multiplication of $-I \in \Gamma$ if necessary, we reach  
\begin{equation}
{\bf{N}} \gamma = 
\begin{pmatrix}
* & * \\ 0 & x
\end{pmatrix}, \qquad \gamma \in \Gamma,
\end{equation}
where $*$ denotes an unspecified integer. 
The above $\Gamma = SL(2,\mathbb{Z})$ multiplications do not change the determinant. Thus, we can write
\begin{equation}
{\bf{N}} \gamma = 
\begin{pmatrix}
D/x & k \\
0 & x
\end{pmatrix}, \qquad k \in \{ 0, 1, \cdots, D/x-1 \}.
\end{equation}
It can be easily seen that if and only if $k \equiv l,\ ({\rm mod}D/x)$,
the following two matrices
\begin{equation}
\begin{pmatrix}
D/x & k \\ 0 & x
\end{pmatrix},
\begin{pmatrix}
D/x & l \\ 0 & x
\end{pmatrix},
\end{equation}
are in the same Type. Consequently, we have verified that intersection matrices $\bf{N}$ are classified into Types. The number of Types is given by 
\begin{equation}
    \sum_{x} \frac{D}{x} ,
\end{equation}
which is nothing but the sum of all positive divisors of $D={\rm det}\bf{N}$. $\Box$

This way of classification into Types is helpful.
This can be seen when we plot $\vec{j}$ of zero-modes on $(j_1, j_2)$ plane. Each Type corresponds to a distinct alignment 
as $j_2=kj_1$ for all of the three vectors $\vec{j}$, where $k$ is constant in a model.
We refer to the $k$ as the gradient.
This shows that Types we defined in Eq.$(\ref{eq: Type})$ is consistent with Ref.\cite{Abe:2014nla}. There, types were defined in terms of the mod 3 structure of $\bf{N}$ without reference to the group theoretical approach we took. 
Types are shown in Table \ref{tab:D3}.
We define ${\bf{M}}_{(3)}$ by
\begin{equation}
{\bf{M}}_{(3)} = 
 \begin{pmatrix}
 3 & 0 \\
 0 & 1
 \end{pmatrix}.
\end{equation}

\begin{table}[H]
\begin{center}
\begin{tabular}{|c|c|c|c|} \hline
Type  & ${\bf{N}}$ & gradient \\ \hline
I & $ \left\{ {\bf{M}}_{(3)} \Gamma \right\}$ & 0  \\ \hline
II & $\left\{ T{\bf{M}}_{(3)} \Gamma \right\}$ & -1 \\ \hline
III& $\left\{ T^2{\bf{M}}_{(3)} \Gamma \right\}$ & -2 \\ \hline
IV& $\left\{ S{\bf{M}}_{(3)} \Gamma \right\}$ & $\infty$ \\ \hline
\end{tabular}
\end{center}
\caption{Types for $D=3$.}
\label{tab:D3}
\end{table}

The alignment of $\vec{j}$ is shown below. 

Type I :
\begin{gather}
\vec{j} = (0,0), (1/3 , 0), (2/3 , 0) \hspace{90pt} \begin{minipage}{0.2\textwidth}
\setlength{\unitlength}{0.6mm}
\begin{center} 
  \includegraphics[width=4cm]{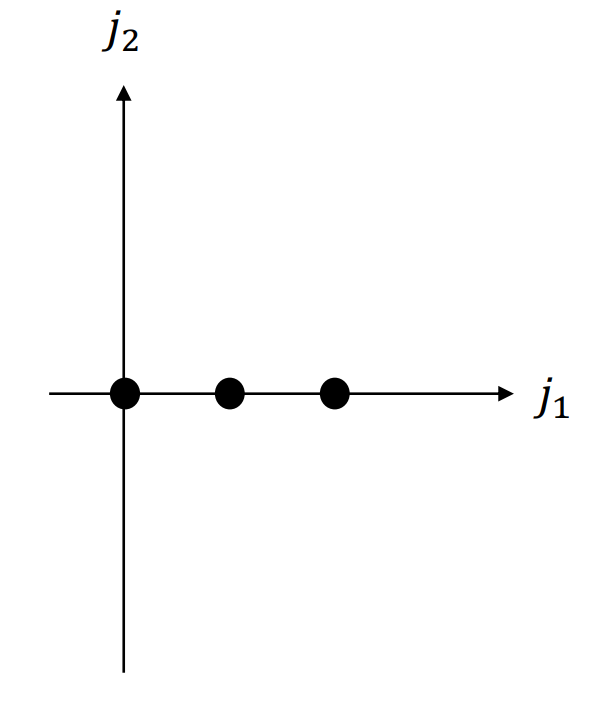}
\end{center}
\end{minipage}\label{type1}
\end{gather}

Type II :
\begin{gather}
\vec{j} = (0,0), (1/3 , -1/3), (2/3 , -2/3) \hspace{40pt} \begin{minipage}{0.2\textwidth}
\setlength{\unitlength}{0.6mm}
\begin{center}
  \includegraphics[width=4cm]{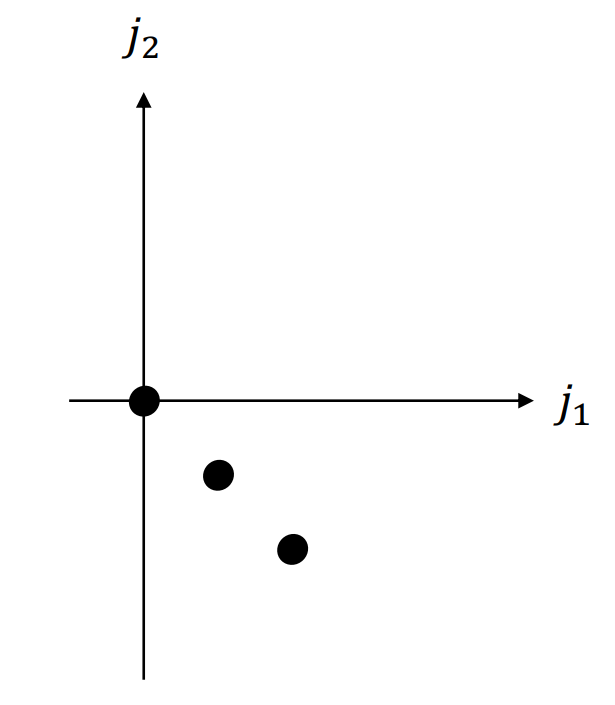}
\end{center}
\end{minipage}\label{type2}
\end{gather}

Type III :
\begin{gather}
 \vec{j} = (0,0), (1/3 , -2/3), (2/3 , -4/3) \hspace{47pt} \begin{minipage}{0.2\textwidth}
\setlength{\unitlength}{0.6mm}
\begin{center} 
  \includegraphics[width=4cm]{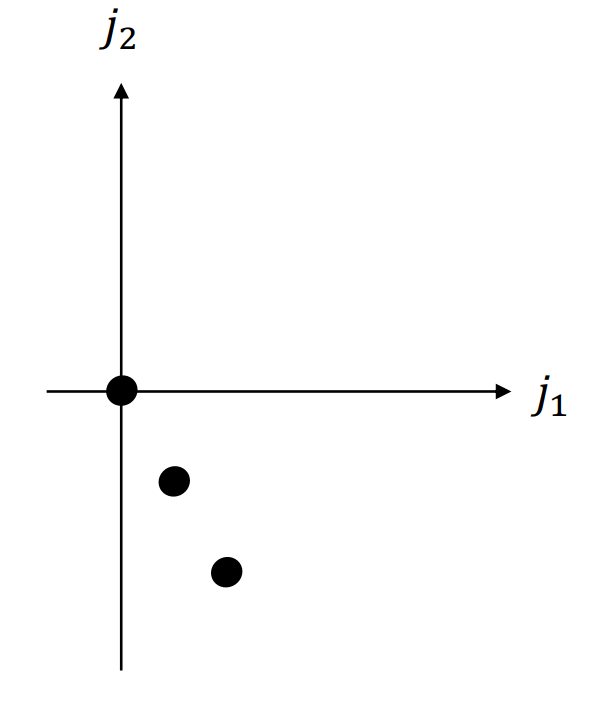}
 \end{center}
\end{minipage}
\end{gather}

Type IV :
\begin{gather}
\label{eq: TypeIV}
\vec{j} = (0,0), (0 , 1/3), (0 , 2/3) \hspace{90pt} \begin{minipage}{0.2\textwidth}
\begin{center}
 \includegraphics[width=4cm]{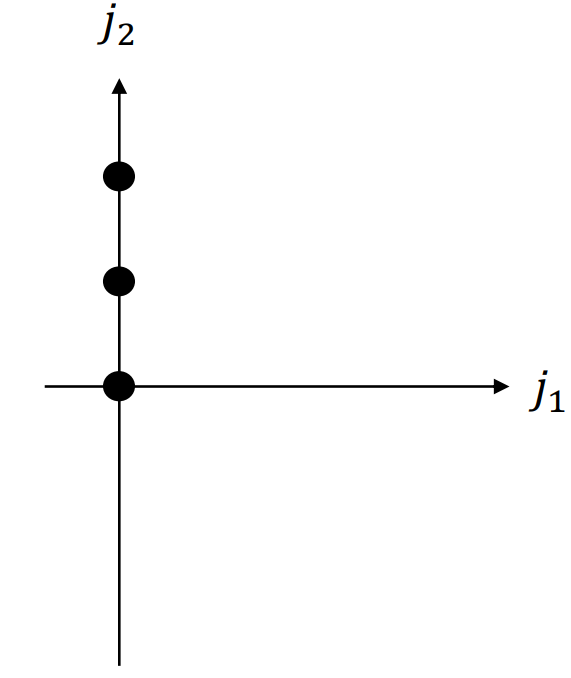}
\end{center}
\end{minipage}
\end{gather}

\subsection{Type and alignment of zero-modes}
We saw that zero-modes are labelled by two dimensional real vectors $\vec{j}$. Under the intersection matrix ${\bf{N}}_1$, $\vec{j}$ must satisfy 
\begin{equation}
\label{eq: j_condition}
\vec{j} \cdot {\bf{N}}_1 = \vec{m} \in \mathbb{Z}^2.
\end{equation}
Suppose that ${\bf{N}}_1$ and ${\bf{N}}_2$ are in the same Type, namely ${\bf{N}}_2 = {\bf{N}}_1 \gamma,\ \gamma \in \Gamma$.
We multiply $\gamma$ from the right to both sides,
\begin{equation}
\vec{j} \cdot {\bf{N}}_2 = \vec{m} \gamma \in \mathbb{Z}^2.
\end{equation}
This shows that all $\vec{j}$ yielded in the ${\bf{N}}_1$ case are also present in the ${\bf{N}}_2$ case and vice versa. Consequently, intersection matrices in the same Type produce the same $\vec{j}$.  

\paragraph{Gradient} \ \\
Referring to Table $\ref{tab:D3}$ and Eqs.$(\ref{type1})$ - $(\ref{eq: TypeIV})$, it seems that acting $T$ to the left of $\bf{N}$ decreases the corresponding gradient of the alignment by 1. On the other hand, acting $S$ rotates the alignment by $\pi/2$. Here we clarify these observations. Suppose zero-modes  under $\bf{N}$ are aligned with gradient $k$. Then we can write 
\begin{equation}
\vec{j} = (j_1, j_2) = (j_1, k j_1). 
\end{equation}
First, let us consider the action of $T$. We can rewrite the condition Eq.($\ref{eq: j_condition}$) as
\begin{equation}
(\vec{j} \cdot T^{-1} ) \cdot  T {\bf{N}} = \vec{m} \in \mathbb{Z}^2.
\end{equation}
Thus, zero-modes labelled by $\vec{j} \cdot T^{-1}$ are yielded under the intersection matrix $T {\bf{N}}$,
\begin{equation}
\vec{j} \cdot T^{-1} = 
(j_1, k j_1)
 \begin{pmatrix}
 1 & -1 \\
 0 & 1 
 \end{pmatrix} = 
 (j_1 , (k-1)j_1).
\end{equation}
This shows that zero-modes under $T{\bf{N}}$ are aligned with gradient $k-1$. Next, let us consider the action of $S$ in the same way. Zero-modes under $S {\bf{N}}$ are written as
\begin{equation}
\vec{j} \cdot S^{-1} =
(j_1, kj_1) 
 \begin{pmatrix}
 0 & 1\\
 -1 & 0  
 \end{pmatrix} =
 (-kj_1, j_1 ),
\end{equation}
showing that their alignment is perpendicular to that of $\vec{j}$. 

One more comment should be made. There is an arbitrariness in the gradient value by $\mod D$. For example, we could say that Type II in Table $\ref{tab:D3}$ corresponds to gradient $2$ instead of $-1$. To understand this, let us consider
\begin{equation}
    (\vec{j}\cdot T^{-D}) \cdot T^{D} {\bf{N}} = \vec{m} \in \mathbb{Z}^2.
\end{equation}
It is easily verified that $T^{D} {\bf{N}}$ and $\bf{N}$ are in the same Type, therefore the gradient is fixed up to $D$. This also explains why we do not get $\{ T^3  {\bf{M}}_{(3)} \Gamma \}$ as an independent Type.

\subsection{Consistency with SUSY}
We have classified all ${\bf{N}}$ whose determinant is 3 without careful considerations of the SUSY condition. We would like to clarify whether it is possible to find $\Omega$ satisfying  $({\bf{N}}\Omega)^T = {\bf{N}} \Omega$ for any given ${\bf{N}}$. In fact, we can always find such $\Omega$ as it will be shown in this section.

 However, if we further demand $\Omega^T = \Omega$, we have severe restrictions. Set of such symmetric $\Omega$ are referred to as Siegel upper half space $\mathcal{H}_{g=2}$. When $\Omega \in \mathcal{H}_2$, $T^4$ is known as Jacobian variety \cite{Mumford:1983}. 
In this case, the SUSY condition is not always satisfied and we study this point as well in detail. This analysis may be useful when considering modular transformations described by $Sp(2g,\mathbb{Z})$. 

\subsubsection{Demanding $\Omega^T = \Omega$}
Let us first analyze the SUSY condition under the requirement $\Omega \in \mathcal{H}_2$. Let us parametrize $\bf{N}$ and $\Omega$ by
\begin{equation}
{\bf{N}} = 
 \begin{pmatrix}
 n_{11} & n_{12} \\
 n_{21} & n_{22}
 \end{pmatrix},\quad
\Omega = 
 \begin{pmatrix}
 \omega_1 & \omega_2 \\
 \omega_2 & \omega_4
 \end{pmatrix}.
\end{equation}
Then the SUSY condition is equivalent to 
\begin{equation}
\label{eq: SUSY_}
(n_{11} - n_{22}) \omega_2 = n_{21} \omega_1 - n_{12} \omega_4 .
\end{equation}

${\rm Im} \Omega$ must be positive definite. Thus, the following two conditions need to be satisfied simultaneously,
\begin{align}
\label{eq: Im_omega1}
{\rm Tr} ({\rm Im} \Omega) > 0,\\
\label{eq: Im_omega2}
{\rm det} ({\rm Im} \Omega) > 0.
\end{align}
We discuss two cases with $({\rm I})\ n_{11} \neq n_{22}$ and $({\rm II})\ n_{11} = n_{22}$ separately 
in what follows.

\paragraph{$({\rm I})\ n_{11} \neq n_{22}:$}\ \\
From Eq.$(\ref{eq: SUSY_})$, we can write 
\begin{equation}
{\rm Im} \Omega = 
 \begin{pmatrix}
 {\rm Im} \omega_1 & \frac{1}{n_{11} - n_{22}} (n_{21} {\rm Im}\omega_1 - n_{12} {\rm Im}\omega_4)  \\
\frac{1}{n_{11} - n_{22}} (n_{21} {\rm Im}\omega_1 - n_{12} {\rm Im}\omega_4) & {\rm Im} \omega_4
 \end{pmatrix}.
\end{equation}
This allows us to write Eqs.$(\ref{eq: Im_omega1})$ and $(\ref{eq: Im_omega2})$ as
\begin{align}
\label{eq: trace_cond}
{\rm Im} \omega_1 &+ {\rm Im} \omega_4 > 0, \\
\label{eq: det_cond}
({\rm Im} \omega_1) ({\rm Im} \omega_4) - &\Delta^2 (n_{21} {\rm Im} \omega_1 - n_{12} {\rm Im} \omega_4 )^2 > 0, 
\end{align}
where 
\begin{equation}
0 < \Delta^2 \equiv \left( \frac{1}{n_{11} - n_{22}} \right)^2  \leq 1.
\end{equation}
We notice that ${\rm Im}\omega_1 > 0,\ {\rm Im}\omega_4 > 0$ must hold. Then, Eq.($\ref{eq: trace_cond}$) is automatically satisfied, and we focus on Eq.($\ref{eq: det_cond}$). Dividing both sides by $({\rm Im}\omega_1)^2$ gives us    
\begin{equation}
- \Delta^2 n_{12}^2 x^2 + (2 \Delta^2 n_{12} n_{21} + 1) x - \Delta^2 n_{21}^2 > 0,
\end{equation}
where $x = {{\rm Im} \omega_4 }/{ {\rm Im} \omega_1} (> 0)$. If $n_{12} = 0$, we get $x > \Delta^2 n_{21}^2$ showing the region of allowed $\Omega \in \mathcal{H}_2$. If $n_{12} \neq 0$, we can further divide both sides by $- \Delta^2 n_{12}^2$ to get 
\begin{equation}
\label{eq: omega_region0}
x^2 - \left( \frac{2 n_{21}}{n_{12}} + \frac{1}{\Delta^2 n_{12}^2} \right)x + \left( \frac{n_{21}}{n_{12}} \right)^2 < 0.
\end{equation}
Thus, the allowed region of $\Omega$ is given by the intersection of $x$ bounded by Eq.($\ref{eq: omega_region0}$) and $x > 0$. In order to get the non-vanishing intersection, we need
\begin{equation}
\label{eq: final_cond}
(n_{11} - n_{22})^2 \geq 2 n_{12}^2 \left( \left| \frac{n_{21}}{n_{12}} \right| - \frac{n_{21}}{n_{12}}\right).
\end{equation}
If $n_{12} n_{21} > 0$ or $n_{21}=0$, Eq.($\ref{eq: final_cond}$) is always satisfied. On the other hand, if $n_{12} n_{21} < 0$ Eq.($\ref{eq: final_cond}$) is written as
\begin{equation}
(n_{11} - n_{22})^2 \geq - 4 n_{12} n_{21}.
\end{equation}
Unless this inequality is met, there is no $\Omega \in \mathcal{H}_2$ satisfying the SUSY condition.

\paragraph{$({\rm II})\ n_{11} = n_{22}:$}\ \\
The SUSY condition is simplified to 
\begin{equation}
n_{21} \omega_1 = n_{12} \omega_4.
\end{equation}
Then, it is easy to show the following results. If ${\bf{N}}$ is of the form  
\begin{equation}
n_{12} n_{21} < 0,\quad  {\rm or}\quad n_{12}=0,\ n_{21} \neq 0,\quad  {\rm or}\quad  n_{12} \neq 0,\ n_{21} = 0,
\end{equation} 
there is no $\Omega \in \mathcal{H}_2$ satisfying the SUSY condition. If $n_{12}n_{21}>0$, $\Omega$ is bounded by
\begin{equation}
{\rm Im} \omega_1 > 0,\quad \frac{|{\rm Im} \omega_2|}{{\rm Im} \omega_1} < \sqrt{\frac{n_{21}}{{n}_{12}}}.
\end{equation}
If $n_{12}=n_{21}=0$, there is no restriction on $\Omega \in \mathcal{H}_2$.

\subsubsection{Not demanding $\Omega^T = \Omega$}
Let us parametrize the complex structure $\Omega$ as
\begin{equation}
\Omega = 
 \begin{pmatrix}
 \omega_1 & \omega_2 \\
 \omega_3 & \omega_4
 \end{pmatrix}.
\end{equation}
The SUSY condition is equivalent to 
\begin{equation}
n_{11} \omega_2 + n_{12} \omega_4 = n_{21} \omega_1 + n_{22} \omega_3.
\end{equation}
We discuss two cases with $({\rm I})\ n_{22} \neq 0$ and $({\rm II})\ n_{22} = 0$ separately 
in what follows.

\paragraph{$({\rm I})\ n_{22} \neq 0\ :$}\ \\ 
We can write
\begin{equation}
{\rm Im} \Omega = 
 \begin{pmatrix}
 {\rm Im} \omega_1 & {\rm Im}\omega_2 \\
 \frac{1}{n_{22}} (n_{11} {\rm Im}\omega_2 + n_{12} {\rm Im}\omega_4 - n_{21} {\rm Im}\omega_1) & {\rm Im} \omega_4
 \end{pmatrix}.
\end{equation}
Then the condition ${\rm Im} \Omega > 0$ is equivalent to the following two inequalities:
\begin{align}
\begin{aligned}
{\rm Im} \omega_1 +& {\rm Im} \omega_4 > 0, \\
( {\rm Im}\omega_1) ( {\rm Im}\omega_4) -& \frac{1}{n_{22}}  {\rm Im}\omega_2 (n_{11}  {\rm Im}\omega_2 + n_{12} {\rm Im}\omega_4 - n_{21}  {\rm Im}\omega_1) > 0.
\end{aligned}
\end{align}
First, let us consider the case when ${\rm Im}\omega_2 > 0$. Then we get 
\begin{align}
\begin{aligned}
X &+ Y > 0, \\
XY - \frac{1}{n_{22}} &(n_{11} + n_{12} Y - n_{21} X) > 0,
\end{aligned}
\end{align}
where $X = {\rm Im} \omega_1/ {\rm Im} \omega_2$ and $Y = {\rm Im} \omega_4/ {\rm Im} \omega_2$. We can see that $\Omega$ such that $X + Y \gg 0$ and $XY \gg 0$ can satisfy the SUSY condition regardless of ${\bf{N}}$. To be more preceise, the curve given by
\begin{equation}
Y = \frac{1}{(n_{22})^2} \cdot \frac{{\rm det}{\bf{N}}}{X- n_{12}/n_{21}} - \frac{n_{21}}{n_{22}},
\end{equation}
and the line $Y=-X$ become the boundaries of $\Omega$ satisfying the SUSY condition. 
The case with ${\rm Im}\omega_2 < 0$ can be analyzed in the same way, and again we always get an allowed region of $\Omega$. 
The case with ${\rm Im} \omega_2=0$ is simple, and the region corresponds to ${\rm Im}\omega_1>0,\ {\rm Im}\omega_4>0$.

\paragraph{$({\rm II})\ n_{22} = 0\ :$}\ \\ 
The SUSY condition is simplified to 
\begin{equation}
n_{11} \omega_2 + n_{12} \omega_4 = n_{21} \omega_1.
\end{equation}
If $n_{11} \neq 0$, we obtain
\begin{equation}
{\rm Im} \Omega = 
 \begin{pmatrix}
 {\rm Im} \omega_1 & \frac{1}{n_{11}} (n_{21} {\rm Im} \omega_1 - n_{12} {\rm Im} \omega_4) \\
 {\rm Im} \omega_3 & {\rm Im} \omega_4
 \end{pmatrix}.
\end{equation}
The condition ${\rm Im}\Omega > 0$ is equivalent to the following two inequalities:
\begin{align}
\begin{aligned}
{\rm Im} \omega_1 &+ {\rm Im} \omega_4 > 0, \\
({\rm Im} \omega_1) ({\rm Im} \omega_4) - &\frac{1}{n_{11}} {\rm Im} \omega_3 \cdot (n_{21} {\rm Im} \omega_1 - n_{12} {\rm Im} \omega_4) > 0.
\end{aligned}
\end{align}
We can always find $\Omega$ satisfying the above conditions.
If $n_{11} = 0$, $\Omega$ is bounded by the following two inequalities:
\begin{align}
\begin{aligned}
\frac{n_{12} + n_{21}}{n_{12}} ({\rm Im} \omega_1) &> 0, \\
\frac{n_{21}}{n_{12}} ({\rm Im} \omega_1)^2 - ({\rm Im} \omega_3)&({\rm Im} \omega_4) > 0.
\end{aligned}
\end{align}
We can always find $\Omega$ satisfying the above conditions.

\section{$T^4/Z_2$}

We constructed $Z_2$ even wave functions on $T^4/Z_2$ by the sum of two zero-modes on $T^4$ as shown in Eq.$(\ref{eq: Z_2_zero})$. For some particular $\vec{j}$, the two modes are identical,
\begin{equation}
\psi^{(\vec{j}+\vec{\alpha} {\bf{N}}^{-1}, \vec{\beta})}(\vec{z}, \Omega) = \psi^{(\vec{j}+\vec{\alpha} {\bf{N}}^{-1}, \vec{\beta})}(-\vec{z}, \Omega).
\end{equation}
Thus, the corresponding $Z_2$ even mode $\psi_{T^4/Z_2,+}$ is written by a single wave function $\psi$ on $T^4$. We denote the number of such invariant modes by $C_+$. Similarly, for some particular $\vec{j}$ we have
\begin{equation}
\psi^{(\vec{j}+\vec{\alpha} {\bf{N}}^{-1}, \vec{\beta})}(\vec{z}, \Omega) =- \psi^{(\vec{j}+\vec{\alpha} {\bf{N}}^{-1}, \vec{\beta})}(-\vec{z}, \Omega).
\end{equation}
Then from Eq.(\ref{eq: Z_2_zero_odd}), we notice that the corresponding $Z_2$ odd mode $\psi_{T^4/Z_2,-}$ is written by a single wave function $\psi$ on $T^4$. We denote the number of such modes by $C_-$. 
The degeneracies of the zero-modes for $Z_2$ even and odd parities are given by the following formulae respectively
\begin{align}
\begin{aligned}
\label{eq: generation_counting}
N_{\rm even} &= \frac{D}{2} + \frac{1}{2} (C_+ - C_-), \\
N_{\rm odd} &= \frac{D}{2} - \frac{1}{2} (C_+ - C_-). 
\end{aligned}
\end{align}
This motivates us to study how $C_+$ and $C_-$ are determined. As it will be shown later, they depend on Scherk-Schwarz phases and the mod 2 structure of $\bf{N}$. The mod 2 structure is simply given by taking mod 2 of each matrix element of $\bf{N}$.
For example, if we have
\begin{equation}
   \bf{N}= 
   \begin{pmatrix}
    5 & 1 \\
    0 & 1
    \end{pmatrix},
\end{equation}
the corresponding mod 2 structure is 
\begin{equation}
   \begin{pmatrix}
    1 & 1 \\
    0 & 1
    \end{pmatrix},
\end{equation}
and we often denote this by writing $(n_{11}, n_{12}, n_{21}, n_{22})\equiv(1,1,0,1),\ ({\rm mod} 2)$. Table $\ref{tb:alpha0}$ summarizes the result when $\vec{\alpha}=(0,0)$. 
\begin{table}[H]
  \centering
  \begin{tabular}{|c||c|c|c|c|} \hline 
  \begin{tabular}{c} mod 2 structure\\ $(n_{11}, n_{12}, n_{21}, n_{22})$ \end{tabular} & $\vec{\beta} = (0,0)$ &  
    $\vec{\beta}=(0,\frac{1}{2})$ &  
    $\vec{\beta}=(\frac{1}{2},0)$ & 
    $\vec{\beta}=(\frac{1}{2},\frac{1}{2})$ 
   \\ \hline 
    $(0,0,0,0)$ & \begin{tabular}{c} $C_+ = 4$ \\ $C_- = 0 $ \end{tabular} & \multicolumn{3}{|c|}{\begin{tabular}{c} $C_+ = 2$ \\ $C_- = 2$ \end{tabular} } 
    \\ \cline{1-5} 
    $\begin{tabular}{c} $(0,0,0,1)$ \\ $(0,0,1,0)$\\ $(0,0,1,1)$ \end{tabular} $ & \multirow{8}{*}{\begin{tabular}{c} $C_+ = 2$ \\ $C_- = 0 $ \end{tabular}}   &   \begin{tabular}{c} $C_+ = 2$ \\ $C_- = 0 $ \end{tabular}  &  \begin{tabular}{c} $C_+ = 1$ \\ $C_- = 1 $ \end{tabular} & \begin{tabular}{c} $C_+ = 1$ \\ $C_- = 1 $ \end{tabular} \\ \cline{1-1}\cline{3-5} 
    $\begin{tabular}{c} $(0,1,0,0)$ \\ $(1,0,0,0)$\\ $(1,1,0,0)$ \end{tabular}$ &    &  \begin{tabular}{c} $C_+ = 1$ \\ $C_- = 1 $ \end{tabular} &   \begin{tabular}{c} $C_+ = 2$ \\ $C_- = 0 $ \end{tabular} &  \begin{tabular}{c} $C_+ = 1$ \\ $C_- = 1 $ \end{tabular}   \\ \cline{1-1}\cline{3-5} 
    $\begin{tabular}{c} $(0,1,0,1)$ \\ $(1,0,1,0)$\\ $(1,1,1,1)$ \end{tabular}$ &  & \begin{tabular}{c} $C_+ = 1$ \\ $C_- = 1 $ \end{tabular} & \begin{tabular}{c} $C_+ = 1$ \\ $C_- = 1 $ \end{tabular} & \begin{tabular}{c} $C_+ = 2$ \\ $C_- = 0 $ \end{tabular}      \\ \hline
     $\begin{tabular}{c} $(0,1,1,0)$ \\ $(1,0,0,1)$\\ $(0,1,1,1)$ \\ $(1,0,1,1)$ \\ $(1,1,0,1)$\\ $(1,1,1,0)$ \end{tabular}$ &  \multicolumn{4}{|c|}{\begin{tabular}{c} $C_+ = 1$ \\ $C_- = 0 $ \end{tabular} }  
 \\ \hline \end{tabular} 
  \caption{$C_{\pm}$ when $\vec{\alpha} = (0,0)$}
  \label{tb:alpha0}
\end{table}

\paragraph{Derivation of Table $\ref{tb:alpha0}$}\ \\
First, let us look at the simplest setup where Scherk-Schwarz phases are vanishing ($\vec{\alpha} = \vec{\beta} = \vec{0}$). We saw that wave functions with $\vec{j} = (0,0), (0, \frac{1}{2}), (\frac{1}{2},0), (\frac{1}{2}, \frac{1}{2})$, are $Z_2$  invariant modes. Thus, $C_+$ can take at most 4. On the other hand, $C_-$ is always 0. We investigate the relationship between intersection matrices $\bf{N}$ and $C_+$ values in detail. Let us look at
\begin{equation}
\vec{j} \cdot {\bf N} = 
(j_1, j_2) \begin{pmatrix} n_{11} & n_{12} \\ n_{21} & n_{22} \end{pmatrix}
= (n_{11} j_1 + n_{21} j_2 , n_{12} j_1 + n_{22} j_2). 
\end{equation}
We notice $\vec{j} = (0,0)$ always appears as a zero-mode state because $\vec{j} \cdot {\bf{N}} = \vec{0} \in \mathbb{Z}^2$. On the other hand, the remaining 3 invariant modes only exist when $\bf{N}$ has a certain structure. Firstly, $(0, \frac{1}{2})$ appears when $n_{21} \equiv 0, n_{22} \equiv 0\ ({\rm mod}2)$. Secondly, $(\frac{1}{2},0)$ appears when $n_{11} \equiv 0, n_{12} \equiv 0\ ({\rm mod}2)$. Lastly, $(\frac{1}{2},\frac{1}{2})$ appears when $n_{11}+n_{21} \equiv 0, n_{12} + n_{22} \equiv 0\ ({\rm mod}2)$. The relationship between $C_+$ and the mod 2 structure of $\bf{N}$ is summarized below. 
We have 16 different patterns of mod 2 structure in $\bf{N}$. Only one of them corresponds to the $C_+=4$ case,
\begin{align}
\label{eq: C+4}
C_+ = 4: \quad {\bf{N}} \equiv \begin{pmatrix} 0 & 0 \\ 0 & 0 \end{pmatrix}\ ({\rm mod\ 2})  .
\end{align}
There is no case with $C_+=3$. There are 9 patterns in the $C_+=2$ case as shown below,    
\begin{align}
C_+ = 2 :  & \nonumber  \\
\label{eq: C+2}
&\vec{j} =(0,0), (0,\frac{1}{2}); \quad{\bf{N}} \equiv 
\begin{pmatrix} 0 & 1 \\ 0 & 0 \end{pmatrix},
\begin{pmatrix} 1 & 0 \\ 0 & 0 \end{pmatrix},
\begin{pmatrix} 1 & 1 \\ 0 & 0 \end{pmatrix}\ ({\rm mod\ 2}) , \\
&\vec{j} = (0,0), (\frac{1}{2},0); \quad{\bf{N}} \equiv 
\begin{pmatrix} 0 & 0 \\ 0 & 1 \end{pmatrix},
\begin{pmatrix} 0 & 0 \\ 1 & 0 \end{pmatrix},
\begin{pmatrix} 0 & 0 \\ 1 & 1 \end{pmatrix}\ ({\rm mod\ 2}) , \\
&\vec{j} = (0,0), (\frac{1}{2},\frac{1}{2}); \quad{\bf{N}} \equiv 
\begin{pmatrix} 0 & 1 \\ 0 & 1 \end{pmatrix},
\begin{pmatrix} 1 & 0 \\ 1 & 0 \end{pmatrix},
\begin{pmatrix} 1 & 1 \\ 1 & 1 \end{pmatrix}\ ({\rm mod\ 2})  .
\end{align}

The remaining 6 patterns of the mod 2 structure correspond to the case $C_+=1$,
\begin{align}
\begin{aligned}
\label{eq: C+1}
C_+ = 1 :  &  \\
&\vec{j}=(0,0); \quad{\bf{N}} \equiv 
\begin{pmatrix} 0 & 1 \\ 1 & 0 \end{pmatrix},
\begin{pmatrix} 1 & 0 \\ 0 & 1 \end{pmatrix},
\begin{pmatrix} 0 & 1 \\ 1 & 1 \end{pmatrix},\\
&  \hspace{28.8mm} \ \ 
\begin{pmatrix} 1 & 0 \\ 1 & 1 \end{pmatrix},
\begin{pmatrix} 1 & 1 \\ 0 & 1 \end{pmatrix},
\begin{pmatrix} 1 & 1 \\ 1 & 0 \end{pmatrix}
\ ({\rm mod\ 2}).\quad  
\end{aligned}
\end{align}
Next, we consider non-vanishing Scherk-Schwarz phases. Let us take $\vec{\beta} = (0, \frac{1}{2})$ first. From Eq.($\ref{eq: Z2_wave_flip}$) we can write
\begin{equation}
\psi^{(\vec{j}, \vec{\beta}=(0,\frac{1}{2}))} (- \vec{z}, \Omega) = e^{-2\pi i j_2} \psi^{(\vec{e}-\vec{j}, \vec{\beta}=(0,\frac{1}{2}))} ( \vec{z}, \Omega).
\end{equation}
This shows that zero-modes $(j_1, j_2)=(0, \frac{1}{2}),\ (\frac{1}{2}, \frac{1}{2})$ are no longer $Z_2$ invariant modes, but are $Z_2$ odd modes by themselves. The other two modes $(0,0)$ and $(\frac{1}{2},0)$ are still $Z_2$ invariant. Hence, we just need to make simple modifications to the results Eqs.$(\ref{eq: C+4})- (\ref{eq: C+1})$. For example, the 3 patterns of the mod 2 structure in Eq.($\ref{eq: C+2}$) now correspond to the $C_+=C_-=1$ case. 
Similarly, we can treat $\vec{\beta} = (\frac{1}{2}, 0)$ and $\vec{\beta} = (\frac{1}{2}, \frac{1}{2})$ cases. In the former case, $(j_1, j_2)=(\frac{1}{2},0),\ (\frac{1}{2}, \frac{1}{2})$ are self-odd modes whereas in the latter case, $(j_1, j_2)=(0, \frac{1}{2}),\ (\frac{1}{2}, 0)$ are.   $\Box$

Table $\ref{tb:alpha1}$ summarizes the result when $\vec{\alpha}$ is non-vanishing. 
\begin{table}[H]
  \centering
  \begin{tabular}{|c|c|c||c|c|c|c|} \hline 
  \multicolumn{3}{|c||}{mod 2 structure $(n_{11}, n_{12}, n_{21}, n_{22})$} & \multirow{2}{*}{$\vec{\beta} = (0,0)$} &  
    \multirow{2}{*}{$\vec{\beta}=(0,\frac{1}{2})$} &  
    \multirow{2}{*}{$\vec{\beta}=(\frac{1}{2},0)$} & 
    \multirow{2}{*}{$\vec{\beta}=(\frac{1}{2},\frac{1}{2})$} 
   \\ \cline{1-3} 
   $\vec{\alpha}=(0,\frac{1}{2})$ &  $\vec{\alpha}=(\frac{1}{2},0)$  &  $\vec{\alpha}=(\frac{1}{2},\frac{1}{2})$ & & & & \\ \hline
    $(0,0,0,1)$ & $(0,0,1,0)$ & $(0,0,1,1)$& \multirow{4}{*}{\begin{tabular}{c} $C_+ = 2$ \\ $C_- = 0 $ \end{tabular} }& \begin{tabular}{c} $C_+ = 0$ \\ $C_- = 2 $ \end{tabular} & \begin{tabular}{c} $C_+ = 1$ \\ $C_- = 1 $ \end{tabular}  & 
\begin{tabular}{c} $C_+ = 1$ \\ $C_- = 1 $ \end{tabular}
    \\  \cline{1-3} \cline{5-7}
   $(0,1,0,0)$  & $(1,0,0,0)$ & $(1,1,0,0)$ &   & \begin{tabular}{c} $C_+ = 1$ \\ $C_- = 1 $ \end{tabular}& \begin{tabular}{c} $C_+ = 0$ \\ $C_- = 2 $ \end{tabular}  & \begin{tabular}{c} $C_+ = 1$ \\ $C_- = 1 $ \end{tabular} \\ \cline{1-3} \cline{5-7}
  $(0,1,0,1)$   &  $(1,0,1,0)$  & $(1,1,1,1)$ & &  \begin{tabular}{c} $C_+ = 1$ \\ $C_- = 1 $ \end{tabular} &  \begin{tabular}{c} $C_+ = 1$ \\ $C_- = 1 $ \end{tabular}  & \begin{tabular}{c} $C_+ = 0$ \\ $C_- = 2$ \end{tabular}  \\ \hline
    $\begin{tabular}{c} $(1,0,1,1)$ \\ $(1,1,1,0)$ \end{tabular}$ &  $\begin{tabular}{c} $(0,1,1,1)$ \\ $(1,1,0,1)$ \end{tabular}$ & \begin{tabular}{c} $(0,1,1,0)$ \\ $(1,0,0,1)$ \end{tabular} & \multirow{6}{*}{\begin{tabular}{c} $C_+ = 1$ \\ $C_- = 0 $ \end{tabular}   } & \begin{tabular}{c} $C_+ = 0$ \\ $C_- = 1 $ \end{tabular}  &  \begin{tabular}{c} $C_+ = 0$ \\ $C_- = 1 $ \end{tabular}    &  \begin{tabular}{c} $C_+ = 1$ \\ $C_- = 0 $ \end{tabular}     \\ \cline{1-3} \cline{5-7}
     $\begin{tabular}{c} $(1,0,0,1)$ \\ $(1,1,0,1)$ \end{tabular}$ &  $\begin{tabular}{c} $(0,1,1,0)$ \\ $(1,1,1,0)$ \end{tabular}$   &  $\begin{tabular}{c} $(1,0,1,1)$ \\ $(0,1,1,1)$ \end{tabular}$  & & \begin{tabular}{c} $C_+ = 0$ \\ $C_- = 1 $ \end{tabular} & \begin{tabular}{c} $C_+ = 1$ \\ $C_- = 0 $ \end{tabular} & \begin{tabular}{c} $C_+ = 0$ \\ $C_- = 1 $ \end{tabular} 
 \\ \cline{1-3} \cline{5-7}
      $\begin{tabular}{c} $(0,1,1,0)$ \\ $(0,1,1,1)$ \end{tabular}$ & $\begin{tabular}{c} $(1,0,0,1)$ \\ $(1,0,1,1)$ \end{tabular}$  & \begin{tabular}{c} $(1,1,1,0)$ \\ $(1,1,0,1)$ \end{tabular} & & \begin{tabular}{c} $C_+ = 1$ \\ $C_- = 0 $ \end{tabular} & \begin{tabular}{c} $C_+ = 0$ \\ $C_- = 1 $ \end{tabular} &
\begin{tabular}{c} $C_+ = 0$ \\ $C_- = 1 $ \end{tabular} 
 \\ \hline
 \begin{tabular}{c} $(0,0,0,0)$ \\ $(0,0,1,0)$ \\ $(1,0,0,0)$ \\ $(1,0,1,0)$ \\ $(0,0,1,1)$ \\ $(1,1,0,0)$ \\ $(1,1,1,1)$ \end{tabular} &  \begin{tabular}{c} $(0,0,0,0)$ \\ $(0,0,0,1)$ \\ $(0,1,0,0)$ \\ $(0,1,0,1)$ \\ $(0,0,1,1)$ \\ $(1,1,0,0)$ \\ $(1,1,1,1)$ \end{tabular}  &\begin{tabular}{c} $(0,0,0,0)$ \\ $(0,0,1,0)$ \\ $(1,0,0,0)$ \\ $(1,0,1,0)$ \\ $(0,0,0,1)$ \\ $(0,1,0,0)$ \\ $(0,1,0,1)$ \end{tabular}  & \multicolumn{4}{|c|}{\begin{tabular}{c} $C_+ = 0$ \\ $C_- = 0 $ \end{tabular} } \\ \hline
 \end{tabular} 
  \caption{$C_{\pm}$ when $\vec{\alpha} = (0,\frac{1}{2}), (\frac{1}{2},0), (\frac{1}{2},\frac{1}{2})$}
  \label{tb:alpha1}
\end{table}

\paragraph{Derivation of Table $\ref{tb:alpha1}$}\ \\
Here we show how to derive the results shown in Table $\ref{tb:alpha1}$. Let us first consider the situation when Scherk-Schwarz phases are $\vec{\alpha}=(0,\frac{1}{2}), \vec{\beta}=(0,0)$. 
From Eq.($\ref{eq: Z2_wave_flip}$) we can write
\begin{equation}
\psi^{(\vec{J}, \vec{\beta}=\vec{0})} (- \vec{z}, \Omega) = \psi^{(\vec{e}-\vec{J}, \vec{\beta}=\vec{0})} ( \vec{z}, \Omega),\qquad \vec{J} = \vec{j} + (0, \frac{1}{2}) {\bf{N}}^{-1}.
\end{equation}
Thus, $Z_2$ invariant modes are labelled by $\vec{J}=(0,0), (0,\frac{1}{2}), (\frac{1}{2},0), (\frac{1}{2},\frac{1}{2})$. We also notice $C_-=0$ under the current setup. Let us investigate the relationship between intersection matrices $\bf{N}$ and $C_+$ values in detail. 
We can write 
\begin{equation}
\vec{j} \cdot {\bf{N}} = (\vec{J}-(0,1/2){\bf{N}}^{-1}) {\bf{N}} 
= (n_{11} J_1 + n_{21} J_2, n_{12} J_1 + n_{22} J_2 - 1/2).
\end{equation}
It is obvious that $\vec{J}=(0,0)$ never appear. On the other hand, the remaining 3 modes exist when $\bf{N}$ has a certain mod 2 structure. Firstly, $\vec{J}=(0,\frac{1}{2})$ appears when $n_{21} \equiv 0,\ n_{22} \equiv 1, ({\rm mod2})$. Secondly, $\vec{J}=(\frac{1}{2},0)$ appears when $n_{11} \equiv 0,\ n_{12} \equiv 1, ({\rm mod2})$. Lastly, $\vec{J}=(\frac{1}{2},\frac{1}{2})$ appears when $n_{11} + n_{21} \equiv 0,\ n_{12} + n_{22} \equiv 1, ({\rm mod2})$. We can summarize as follows,
\begin{align}
\begin{aligned}
\label{eq: SS_C+2}
C_+ = 2 : &  \\
&\vec{J}=(0,\frac{1}{2}), (\frac{1}{2}, \frac{1}{2}); \quad{\bf{N}} \equiv 
\begin{pmatrix} 0 & 0 \\ 0 & 1 \end{pmatrix},\ ({\rm mod\ 2}),  \\
&\vec{J}=(\frac{1}{2},0), (\frac{1}{2}, \frac{1}{2}); \quad{\bf{N}} \equiv 
\begin{pmatrix} 0 & 1 \\ 0 & 0 \end{pmatrix},\ ({\rm mod\ 2}),\\
&\vec{J}=(0,\frac{1}{2}), (\frac{1}{2},0); \quad{\bf{N}} \equiv 
\begin{pmatrix} 0 & 1 \\ 0 & 1 \end{pmatrix},
\ ({\rm mod\ 2}),
\end{aligned}
\end{align}

\begin{align}
\begin{aligned}
\label{eq: SS_C+1}
C_+ = 1 : &  \\
&\vec{J}=(0,\frac{1}{2}); \quad{\bf{N}} \equiv 
\begin{pmatrix} 1 & 0 \\ 0 &1 \end{pmatrix},
\begin{pmatrix} 1 & 1 \\ 0 &1 \end{pmatrix}, \ ({\rm mod\ 2}),  \\
&\vec{J}=(\frac{1}{2},0); \quad{\bf{N}} \equiv 
\begin{pmatrix} 0 & 1 \\ 1 & 0 \end{pmatrix},
\begin{pmatrix} 0 & 1 \\ 1 & 1 \end{pmatrix},
\ ({\rm mod\ 2}),\\
&\vec{J}=(\frac{1}{2},\frac{1}{2}); \quad{\bf{N}} \equiv 
\begin{pmatrix} 1 & 0 \\ 1 & 1 \end{pmatrix},
\begin{pmatrix} 1 & 1 \\ 1 & 0 \end{pmatrix},
\ ({\rm mod\ 2}). 
\end{aligned}
\end{align}
The remaining 7 patterns of the mod 2 structure correspond to the $C_+=0$ case. 

Next, we consider non-vanishing $\vec{\beta}$. Let us look at the case when Scherk-Schwarz phases are $\vec{\alpha}=(0,\frac{1}{2}).\ \vec{\beta} = (0, \frac{1}{2})$. From Eq.($\ref{eq: Z2_wave_flip}$) we can write
\begin{equation}
\psi^{(\vec{J}, \vec{\beta}=(0,\frac{1}{2}))} (- \vec{z}, \Omega) = e^{-2\pi i J_2} \psi^{(\vec{e}-\vec{J}, \vec{\beta}=(0,\frac{1}{2}))} ( \vec{z}, \Omega).
\end{equation}
This shows that zero-modes $(J_1, J_2)=(0, \frac{1}{2}),\ (\frac{1}{2}, \frac{1}{2})$ are no longer $Z_2$ invariant modes, but are $Z_2$ odd modes by themselves. The remaining zero-mode $(J_1, J_2)=(\frac{1}{2},0)$ is still $Z_2$ invariant. This suffices to complete the Table $\ref{tb:alpha1}$ because other setups with different Scherk-Schwarz phases can be analyzed in the same way. $\Box$

\subsection{Three-generation models of $T^4/Z_2$}
In order to get 3 degenerated zero-modes in the $Z_2$ even sector, $D(={\rm det}\bf{N})$ must satisfy
\begin{equation}
 D = 6 -  (C_+ - C_-) =: D_+.
\end{equation}
In the $Z_2$ odd sector, $D$ must satisfy
\begin{equation}
 D = 6 +  (C_+ - C_-) =: D_-.
\end{equation}
We summarize the conditions to get 3 degeneracy by presenting Tables $\ref{tb:alpha0D}$ and $\ref{tb:alpha1D}$. Given $\vec{\alpha}=(0,0)$, Table $\ref{tb:alpha0D}$ shows the required assignments of mod 2 structure, determinant,$Z_2$ parity, and $\vec{\beta}$. For non-vanishing $\vec{\alpha}$, Table $\ref{tb:alpha1D}$ summarizes the result. Note that when the mod 2 structure is of the form $(0,0,0,0)$, the determinant of $\bf{N}$ is a multiple of 4. This fact explains why we cannot produce three-generation models under the mod 2 structure $(0,0,0,0)$. 
From these tables, we find that we can realize three-generation models when $D=4,5,6,7,8$.
We will study these cases in what follows.
\begin{table}[H]
  \centering
  \begin{tabular}{|c||c|c|c|c|} \hline 
    \begin{tabular}{c} mod 2 structure\\ $(n_{11}, n_{12}, n_{21}, n_{22})$ \end{tabular} & $\vec{\beta} = (0,0)$ &  
    $\vec{\beta}=(0,\frac{1}{2})$ &  
    $\vec{\beta}=(\frac{1}{2},0)$ & 
    $\vec{\beta}=(\frac{1}{2},\frac{1}{2})$ 
   \\ \hline 
    $(0,0,0,0)$ &  \multicolumn{4}{|c|}{ no\ solution} 
    \\ \cline{1-5} 
    $\begin{tabular}{c} $(0,0,0,1)$ \\ $(0,0,1,0)$\\ $(0,0,1,1)$ \end{tabular} $ & \multirow{8}{*}{\begin{tabular}{c} $D_+ = 4$ \\ $D_- = 8 $ \end{tabular}}   &   \begin{tabular}{c} $D_+ = 4$ \\ $D_- = 8 $ \end{tabular}  &  \begin{tabular}{c} $D_+ = 6$ \\ $D_- = 6 $ \end{tabular} & \begin{tabular}{c} $D_+ = 6$ \\ $D_- = 6 $ \end{tabular} \\ \cline{1-1}\cline{3-5} 
    $\begin{tabular}{c} $(0,1,0,0)$ \\ $(1,0,0,0)$\\ $(1,1,0,0)$ \end{tabular}$ &    &  \begin{tabular}{c} $D_+ = 6$ \\ $D_- = 6 $ \end{tabular} &   \begin{tabular}{c} $D_+ = 4$ \\ $D_- = 8 $ \end{tabular} &  \begin{tabular}{c} $D_+ = 6$ \\ $D_- = 6$ \end{tabular}   \\ \cline{1-1}\cline{3-5} 
    $\begin{tabular}{c} $(0,1,0,1)$ \\ $(1,0,1,0)$\\ $(1,1,1,1)$ \end{tabular}$ &  & \begin{tabular}{c} $D_+ = 6$ \\ $D_- = 6 $ \end{tabular} & \begin{tabular}{c} $D_+ = 6$ \\ $D_- = 6 $ \end{tabular} & \begin{tabular}{c} $D_+ = 4$ \\ $D_- = 8 $ \end{tabular}      \\ \hline
     $\begin{tabular}{c} $(0,1,1,0)$ \\ $(1,0,0,1)$\\ $(0,1,1,1)$ \\ $(1,0,1,1)$ \\ $(1,1,0,1)$\\ $(1,1,1,0)$ \end{tabular}$ &  \multicolumn{4}{|c|}{\begin{tabular}{c} $D_+ = 5$ \\ $D_- = 7 $ \end{tabular} }  
 \\ \hline \end{tabular} 
  \caption{$\vec{\alpha} = (0,0)$}
  \label{tb:alpha0D}
\end{table}

\begin{table}[H]
  \centering
  \begin{tabular}{|c|c|c||c|c|c|c|} \hline 
  \multicolumn{3}{|c||}{mod 2 structure $(n_{11}, n_{12}, n_{21}, n_{22})$ } & \multirow{2}{*}{$\vec{\beta} = (0,0)$} &  
    \multirow{2}{*}{$\vec{\beta}=(0,\frac{1}{2})$} &  
    \multirow{2}{*}{$\vec{\beta}=(\frac{1}{2},0)$} & 
    \multirow{2}{*}{$\vec{\beta}=(\frac{1}{2},\frac{1}{2})$} 
   \\ \cline{1-3} 
   $\vec{\alpha}=(0,\frac{1}{2})$ &  $\vec{\alpha}=(\frac{1}{2},0)$  &  $\vec{\alpha}=(\frac{1}{2},\frac{1}{2})$ & & & & \\ \hline
    $(0,0,0,1)$ & $(0,0,1,0)$ & $(0,0,1,1)$& \multirow{4}{*}{\begin{tabular}{c} $D_+ = 4$ \\ $D_- = 8 $ \end{tabular} }& \begin{tabular}{c} $D_+ = 8$ \\ $D_- = 4 $ \end{tabular} & \begin{tabular}{c} $D_+ = 6$ \\ $D_- = 6 $ \end{tabular}  & 
\begin{tabular}{c} $D_+ = 6$ \\ $D_- = 6 $ \end{tabular}
    \\  \cline{1-3} \cline{5-7}
   $(0,1,0,0)$  & $(1,0,0,0)$ & $(1,1,0,0)$ &   & \begin{tabular}{c} $D_+ = 6$ \\ $D_- = 6 $ \end{tabular}& \begin{tabular}{c} $D_+ = 8$ \\ $D_- = 4 $ \end{tabular}  & \begin{tabular}{c} $D_+ = 6$ \\ $D_- = 6 $ \end{tabular} \\ \cline{1-3} \cline{5-7}
  $(0,1,0,1)$   &  $(1,0,1,0)$  & $(1,1,1,1)$ & &  \begin{tabular}{c} $D_+ = 6$ \\ $D_- = 6 $ \end{tabular} &  \begin{tabular}{c} $D_+ = 6$ \\ $D_- = 6 $ \end{tabular}  & \begin{tabular}{c} $D_+ = 8$ \\ $D_- = 4$ \end{tabular}  \\ \hline
    $\begin{tabular}{c} $(1,0,1,1)$ \\ $(1,1,1,0)$ \end{tabular}$ &  $\begin{tabular}{c} $(0,1,1,1)$ \\ $(1,1,0,1)$ \end{tabular}$ & \begin{tabular}{c} $(0,1,1,0)$ \\ $(1,0,0,1)$ \end{tabular} & \multirow{6}{*}{\begin{tabular}{c} $D_+ = 5$ \\ $D_- = 7 $ \end{tabular}   } & \begin{tabular}{c} $D_+ = 7$ \\ $D_- = 5 $ \end{tabular}  &  \begin{tabular}{c} $D_+ = 7$ \\ $D_- = 5 $ \end{tabular}    &  \begin{tabular}{c} $D_+ = 5$ \\ $D_- = 7 $ \end{tabular}     \\ \cline{1-3} \cline{5-7}
     $\begin{tabular}{c} $(1,0,0,1)$ \\ $(1,1,0,1)$ \end{tabular}$ &  $\begin{tabular}{c} $(0,1,1,0)$ \\ $(1,1,1,0)$ \end{tabular}$   &  $\begin{tabular}{c} $(1,0,1,1)$ \\ $(0,1,1,1)$ \end{tabular}$  & & \begin{tabular}{c} $D_+ = 7$ \\ $D_- = 5 $ \end{tabular} & \begin{tabular}{c} $D_+ = 5$ \\ $D_- = 7 $ \end{tabular} & \begin{tabular}{c} $D_+ = 7$ \\ $D_- = 5 $ \end{tabular} 
 \\ \cline{1-3} \cline{5-7}
      $\begin{tabular}{c} $(0,1,1,0)$ \\ $(0,1,1,1)$ \end{tabular}$ & $\begin{tabular}{c} $(1,0,0,1)$ \\ $(1,0,1,1)$ \end{tabular}$  & \begin{tabular}{c} $(1,1,1,0)$ \\ $(1,1,0,1)$ \end{tabular} & & \begin{tabular}{c} $D_+ = 5$ \\ $D_- = 7 $ \end{tabular} & \begin{tabular}{c} $D_+ = 7$ \\ $D_- = 5 $ \end{tabular} &
\begin{tabular}{c} $D_+ = 7$ \\ $D_- = 5 $ \end{tabular} 
 \\ \hline
  $(0,0,0,0)$ &  $(0,0,0,0)$ &  $(0,0,0,0)$ & \multicolumn{4}{|c|}{ no\ solution} 
    \\ \hline 
 \begin{tabular}{c} $(0,0,1,0)$ \\ $(1,0,0,0)$ \\ $(1,0,1,0)$ \\ $(0,0,1,1)$ \\ $(1,1,0,0)$ \\ $(1,1,1,1)$ \end{tabular} &  \begin{tabular}{c}  $(0,0,0,1)$ \\ $(0,1,0,0)$ \\ $(0,1,0,1)$ \\ $(0,0,1,1)$ \\ $(1,1,0,0)$ \\ $(1,1,1,1)$ \end{tabular}  &\begin{tabular}{c} $(0,0,1,0)$ \\ $(1,0,0,0)$ \\ $(1,0,1,0)$ \\ $(0,0,0,1)$ \\ $(0,1,0,0)$ \\ $(0,1,0,1)$ \end{tabular}  & \multicolumn{4}{|c|}{\begin{tabular}{c} $D_+ = 6$ \\ $D_- = 6 $ \end{tabular} } \\ \hline
 \end{tabular} 
  \caption{$\vec{\alpha} = (0,\frac{1}{2}), (\frac{1}{2},0), (\frac{1}{2},\frac{1}{2})$}
  \label{tb:alpha1D}
\end{table}
Note that we still get a number of $\bf{N}$ even after both $D$ and mod 2 structure are specified. Therefore, let us classify them. For that purpose, it is helpful to study the relationship between mod 2 structure and group actions on $\bf{N}$. 

\subsection{Mod 2 structure of ${\bf{N}}$}
Here we consider actions of $\gamma \in \Gamma$ on an intersection matrix ${\bf{N}}$ such that 
\begin{equation}
 {\bf{N}}\gamma  \equiv {\bf{N}},\ ({\rm mod}\ 2).
\end{equation}
In other words, we would like to identify the subgroups of $\Gamma$ whose actions conserve the mod 2 structure of $\bf{N}$.
We summarize the results in what follows. 

Following 6 patterns of mod 2 structure are invariant only under the actions of $\Gamma(2) \subset \Gamma$,
\begin{align}
 \begin{pmatrix} 0 & 1 \\ 1 & 0 \end{pmatrix},\ 
 \begin{pmatrix} 1 & 0 \\ 0 & 1 \end{pmatrix},\ 
 \begin{pmatrix} 0 & 1 \\ 1 & 1 \end{pmatrix},\ 
 \begin{pmatrix} 1 & 0 \\ 1 & 1 \end{pmatrix},\ 
 \begin{pmatrix} 1 & 1 \\ 0 & 1 \end{pmatrix},\ 
 \begin{pmatrix} 1 & 1 \\ 1 & 0 \end{pmatrix},
\end{align}
where 
\begin{equation}
\Gamma(2) := \left\{
\begin{pmatrix}
a & b \\
c & d
\end{pmatrix} \in \Gamma : 
\begin{pmatrix}
a & b \\
c & d
\end{pmatrix} \equiv
\begin{pmatrix}
1 & 0 \\
0 & 1
\end{pmatrix}\ ({\rm mod}\ 2)
\right\}.
\end{equation}
$\Gamma(2)$ is called the principal congruence subgroup of level 2. Notice that these six patterns correspond to the case when $D$ is an odd number. 

Following 3 patterns of mod 2 structure are invariant only under the actions of $\Gamma_1(2) \subset \Gamma$,
\begin{equation}
 \begin{pmatrix} 0 & 0 \\ 0 & 1 \end{pmatrix},\ 
 \begin{pmatrix} 0 & 1 \\ 0 & 0 \end{pmatrix},\  
 \begin{pmatrix} 0 & 1 \\ 0 & 1 \end{pmatrix},
\end{equation}
where
\begin{equation}
\Gamma_1(2) := \left\{
\begin{pmatrix}
a & b \\
c & d
\end{pmatrix} \in \Gamma : 
\begin{pmatrix}
a & b \\
c & d
\end{pmatrix} \equiv
\begin{pmatrix}
1 & * \\
0 & 1
\end{pmatrix}\ ({\rm mod}\ 2)
\right\}.
\end{equation}

Following 3 patterns of mod 2 structure are invariant only under the actions of $\Gamma^1 (2) \subset \Gamma$,

\begin{equation}
 \begin{pmatrix} 0 & 0 \\ 1 & 0 \end{pmatrix},\ 
 \begin{pmatrix} 1 & 0 \\ 0 & 0 \end{pmatrix},\  
 \begin{pmatrix} 1 & 0 \\ 1 & 0 \end{pmatrix},
\end{equation}
where
\begin{equation}
\Gamma^1(2) := \left\{
\begin{pmatrix}
a & b \\
c & d
\end{pmatrix} \in \Gamma : 
\begin{pmatrix}
a & b \\
c & d
\end{pmatrix} \equiv
\begin{pmatrix}
1 & 0 \\
* & 1
\end{pmatrix}\ ({\rm mod}\ 2)
\right\}.
\end{equation}

Following 3 patterns of mod 2 structure are invariant only under the action of $\Lambda \subset \Gamma$,
\begin{equation}
 \begin{pmatrix} 0 & 0 \\ 1 & 1 \end{pmatrix},\ 
 \begin{pmatrix} 1 & 1 \\ 0 & 0 \end{pmatrix},\  
 \begin{pmatrix} 1 & 1 \\ 1 & 1 \end{pmatrix},
\end{equation}
where
\begin{equation}
\Lambda := \left\{
\begin{pmatrix}
a & b \\
c & d
\end{pmatrix} \in \Gamma : 
ac \equiv bd \equiv 0 \ ({\rm mod}\ 2)
\right\}.
\end{equation}

$\Gamma_1(2), \Gamma^1(2)$, and $\Lambda$ are known as congruence subgroups of $\Gamma$. 
It is obvious that the mod 2 structure of the form
\begin{equation}
\begin{pmatrix}
0 & 0 \\ 0 & 0
\end{pmatrix},
\end{equation}
is invariant under $\Gamma=SL(2,\mathbb{Z})$. 

\subsection{$D=5$}
Here we show the classifications of $\bf{N}$ when $D=5$. This can be done in two steps. First, we classify them into Types as we did before. Then we further classify each Type in terms of the mod 2 structure. 

When $D=5$, we get $6(=1+5)$ Types as shown in Table \ref{tab:D=5}.
We define ${\bf{M}}_{(5)}$ by
\begin{equation}
{\bf{M}}_{(5)} = 
 \begin{pmatrix}
 5 & 0 \\
 0 & 1
 \end{pmatrix}.
\end{equation}
\begin{table}[H]
\begin{center}
\begin{tabular}{|c|c|c|c|} \hline
Type  & ${\bf{N}}$ & gradient \\ \hline
I & $ \left\{ {\bf{M}}_{(5)} \Gamma \right\}$ & $0$  \\ \hline
II & $\left\{ T{\bf{M}}_{(5)}  \Gamma \right\}$ & $-1$ \\ \hline
III& $\left\{ T^2{\bf{M}}_{(5)}  \Gamma \right\}$ & $-2$ \\ \hline
IV& $\left\{ T^3{\bf{M}}_{(5)}  \Gamma \right\}$ & $-3$ \\ \hline
V & $\left\{ T^4{\bf{M}}_{(5)}  \Gamma \right\}$ & $-4$\\ \hline 
VI & $ \left\{S {\bf{M}}_{(5)}  \Gamma \right\}$ & $\infty$ \\ \hline
\end{tabular}
\end{center}
\caption{Types for $D=5$.}
\label{tab:D=5}
\end{table}

Next, let us consider classifications in terms of the mod 2 structure. First, we concentrate on the set of $\bf{N}$ which belongs to Type I and has mod 2 structure $(1,0,0,1)$. In general, we can write such ${\bf{N}}$ as follows,
\begin{equation}
{\bf{N}} =
 \begin{pmatrix}
 5 m & 10 n \\ * & * 
 \end{pmatrix},\qquad {\rm gcd}(m, 2n)=1,\ m \neq 0.
\end{equation}
We notice that $m$ is an odd number. We act $\Gamma(2)$ to the right of ${\bf{N}}$ as
\begin{equation}
\label{eq: process_a}
 \begin{pmatrix}
 5 m & 10 n \\ * & * 
 \end{pmatrix}
 \begin{pmatrix}
 1 & 2k \\ 0 & 1 
 \end{pmatrix}
 = 
  \begin{pmatrix}
 5 m & 10 (n+km) \\ * & * 
 \end{pmatrix} =: 
  \begin{pmatrix}
 5 m & 10 n' \\ * & * 
 \end{pmatrix},
\end{equation}
where $k \in \mathbb{Z}$. By the appropriate choice of $k$, we can write
\begin{equation}
|n'| < \frac{|m|}{2}.
\end{equation}
We consecutively act another element of $\Gamma(2)$ to the right as 
\begin{equation}
  \begin{pmatrix}
 5 m & 10 n' \\ * & * 
 \end{pmatrix}
   \begin{pmatrix}
 1 & 0  \\ 2l & 1 
 \end{pmatrix}
 = 
    \begin{pmatrix}
 5(m + 4 n' l) & 10 n'  \\ * & * 
 \end{pmatrix}
 =:
     \begin{pmatrix}
 5m' & 10 n'  \\ * & * 
 \end{pmatrix},
\end{equation} 
where $l \in \mathbb{Z}$. If $n' \neq 0$, we can choose $l$ to have
\begin{equation}
|m'| < |2n'|.
\end{equation}
We can repeat the process in Eq.($\ref{eq: process_a}$), and obtain
\begin{equation}
\begin{pmatrix}
 5m' & 10 n'' \\ * & * 
 \end{pmatrix}, \qquad |n''| < |n'|.
\end{equation}
After some iterations, we reach
\begin{equation}
{\bf{N}} \gamma = 
 \begin{pmatrix}
 5 \tilde{m} & 0 \\ * & *
 \end{pmatrix},\qquad \tilde{m} \in \mathbb{Z},\  \gamma \in \Gamma(2).
\end{equation}
${\bf{N}} \gamma$ must have its determinant equal to 5. Thus, we can write 
\begin{equation}
 {\bf{N}} \gamma= \begin{pmatrix}
 \pm 5& 0 \\ * & \pm 1
 \end{pmatrix}.
\end{equation}
Since $-I \in \Gamma(2)$, we can remove the arbitrariness of the overall sign and obtain
\begin{equation}
 {\bf{N}} \gamma= \begin{pmatrix}
 5& 0 \\ 2s & 1
 \end{pmatrix},\qquad s \in \mathbb{Z},\ \gamma \in \Gamma(2).
\end{equation}
Here we used the fact that actions of $\Gamma(2)$ do not change the mod 2 structure. Finally, we act 
\begin{equation}
\begin{pmatrix}
1 & 0 \\ -2s & 1
\end{pmatrix} \in \Gamma(2),
\end{equation}
from the right, and we get 
\begin{equation}
{\bf{N}} = {\bf{M}}_{(5)} \gamma, \quad \gamma \in \Gamma(2).
\end{equation}
It is now easy to write similar expressions for other patterns of the mod 2 structure. For example, set of $\bf{N} \in ({\rm Type\ I})$ with mod 2 structure $(1,1,0,1)$ can be written as $\{ {\bf{M}}_{(5)} T \Gamma(2) \}$. In order to move to different Types, we just need to act $\Gamma$ to the left of ${\bf{N}}$. We summarize the results in Table $\ref{tb: D5mod2}$. Only the representative matrix is shown for each type and mod 2 structure.
For example, by writing ${\bf{M}}_{(5)} T$, we mean the infinite set given by $\{ {\bf{M}}_{(5)} T \Gamma(2) \}$. 

\begin{table}[H]
\setlength{\tabcolsep}{0.8mm} 
\begin{center}
  \begin{tabular}{|c||c|c|c|c|c|c|c|} \hline 
     \begin{tabular}{c}  mod 2 structure   \\  $(n_{11}, n_{12}, n_{21}, n_{22})$  \end{tabular} & Type I &  
    Type II &  
    Type III & 
    Type IV &
    Type V &
    Type VI
   \\ \hline \hline
    $ (1,0,0,1) $& $ {\bf{M}}_{(5)}   $ & $T {\bf{M}}_{(5)}  T  $ & $ T^2 {\bf{M}}_{(5)}  $ & $T^3 {\bf{M}}_{(5)}  T   $ & $T^4 {\bf{M}}_{(5)}   $ &  $S {\bf{M}}_{(5)} S  $ 
     \\ \cline{1-7} 
     
    $ (1,1,0,1) $ &  $ {\bf{M}}_{(5)}  T  $ &   $ T {\bf{M}}_{(5)}   $ & $ T^2 {\bf{M}}_{(5)}  T  $& $ T^3 {\bf{M}}_{(5)}  $ & $ T^4 {\bf{M}}_{(5)}  T $ &  $S {\bf{M}}_{(5)}  ST  $ \\ \cline{1-7} 
    
   $ (0,1,1,0) $ & ${\bf{M}}_{(5)}  S  $   &  $ T {\bf{M}}_{(5)}  TS  $ &   $ T^2 {\bf{M}}_{(5)} S $ &  $ T^3 {\bf{M}}_{(5)}  TS  $ & $ T^4 {\bf{M}}_{(5)}  S  $ & $ S {\bf{M}}_{(5)}  $ \\ \cline{1-7}
   
    $ (1,1,1,0) $ & $ {\bf{M}}_{(5)}  TS $ &  $T {\bf{M}}_{(5)}  S $  & $ T^2 {\bf{M}}_{(5)}  TS $ & $ T^3 {\bf{M}}_{(5)}  S $ & $ T^4 {\bf{M}}_{(5)}  TS $ & $ S {\bf{M}}_{(5)}  STS $ 
  \\ \cline{1-7} 
  
 $ (0,1,1,1) $ & $ {\bf{M}}_{(5)}  ST $& $ T {\bf{M}}_{(5)}  STS $ & $T^2 {\bf{M}}_{(5)} S T  $ & $T^3 {\bf{M}}_{(5)}  STS $ & $ T^4 {\bf{M}}_{(5)}  ST $ & $ S {\bf{M}}_{(5)} T $
 \\ \cline{1-7}
 
 $(1,0,1,1)$ &$ {\bf{M}}_{(5)}  STS  $& $T {\bf{M}}_{(5)}  ST $  & $ T^2 {\bf{M}}_{(5)}  STS$ & $T^3 {\bf{M}}_{(5)}  ST $ & $ T^4 {\bf{M}}_{(5)}  STS $ & $ S {\bf{M}}_{(5)}  TS $ \\
 \cline{1-7}
  \end{tabular} 
  \end{center}
  \caption{Classification by the mod 2 structure for $D=5$}
  \label{tb: D5mod2}
\end{table}
One can understand the reason why we obtain 6 different patterns of mod 2 structure for each Type in the following way. Note that mod 2 structure of $\bf{N}$ with $D= ({\rm odd})$ is invariant only under $\Gamma(2)$ when we consider group actions from the right. Thus, $S_3 \simeq \Gamma/\Gamma(2)$ corresponds to the change of mod 2 structure. Since the order of $S_3$ is 6, there should exist 6 different patterns in each Type. 

\subsection{$D=7$}
Here we show the classifications of $\bf{N}$ when $D=7$. We get $8(=1+7)$ Types as shown in Table \ref{tab:D=7}.
 We define ${\bf{M}}_{(7)}$ by
\begin{equation}
{\bf{M}}_{(7)} = 
 \begin{pmatrix}
 7 & 0 \\
 0 & 1
 \end{pmatrix}.
\end{equation}
\begin{table}[H]
\begin{center}
\begin{tabular}{|c|c|c|c|} \hline
Type  & ${\bf{N}}$ & gradient \\ \hline
I & $ \left\{ {\bf{M}}_{(7)} \Gamma \right\}$ & $0$  \\ \hline
II & $\left\{ T{\bf{M}}_{(7)}  \Gamma \right\}$ & $-1$ \\ \hline
III& $\left\{ T^2{\bf{M}}_{(7)}  \Gamma \right\}$ & $-2$ \\ \hline
IV& $\left\{ T^3{\bf{M}}_{(7)}  \Gamma \right\}$ & $-3$ \\ \hline
V & $\left\{ T^4{\bf{M}}_{(7)}  \Gamma \right\}$ & $-4$\\ \hline 
VI & $ \left\{T^5 {\bf{M}}_{(7)}  \Gamma \right\}$ & $-5$ \\ \hline
VII & $\left\{ T^6{\bf{M}}_{(7)}  \Gamma \right\}$ & $-6$\\ \hline 
VIII & $ \left\{S {\bf{M}}_{(7)}  \Gamma \right\}$ & $\infty$ \\ \hline
\end{tabular}
\end{center}
\caption{Types for $D=7$.}
\label{tab:D=7}
\end{table}

Classifications in terms of the mod 2 structure is straightforward. 
Results are shown in Table \ref{tb:D=7-mod2}.
Type $\rm IV - VI$ are omitted, but it is easy to recover the full results. 
\begin{table}[htbp]
\begin{center}
  \begin{tabular}{|c||c|c|c|c|c|c|c|} \hline 
      \begin{tabular}{c} mod 2 structure\\ $(n_{11}, n_{12}, n_{21}, n_{22})$ \end{tabular} & Type I &  
    Type II &  
    Type III & 
    $\cdots$ &
    Type VII &
    Type VIII
   \\ \hline \hline
    $ (1,0,0,1) $& $ {\bf{M}}_{(7)}   $ & $T {\bf{M}}_{(7)}  T  $ & $ T^2 {\bf{M}}_{(7)}  $ & $ \cdots   $ & $T^6 {\bf{M}}_{(7)}   $ &  $S {\bf{M}}_{(7)} S  $ 
     \\ \cline{1-7} 
     
    $ (1,1,0,1) $ &  $ {\bf{M}}_{(7)}  T  $ &   $ T {\bf{M}}_{(7)}   $ & $ T^2 {\bf{M}}_{(7)}  T  $& $ \cdots$ & $ T^6 {\bf{M}}_{(7)}  T $ &  $S {\bf{M}}_{(7)}  ST  $ \\ \cline{1-7} 
    
   $ (0,1,1,0) $ & ${\bf{M}}_{(7)}  S  $   &  $ T {\bf{M}}_{(7)}  TS  $ &   $ T^2 {\bf{M}}_{(7)} S $ &  $ \cdots  $ & $ T^6 {\bf{M}}_{(7)}  S  $ & $ S {\bf{M}}_{(7)}  $ \\ \cline{1-7}
   
    $ (1,1,1,0) $ & $ {\bf{M}}_{(7)}  TS $ &  $T {\bf{M}}_{(7)}  S $  & $ T^2 {\bf{M}}_{(7)}  TS $ & $ \cdots$ & $ T^6 {\bf{M}}_{(5)}  TS $ & $ S {\bf{M}}_{(7)}  STS $ 
  \\ \cline{1-7} 
  
 $ (0,1,1,1) $ & $ {\bf{M}}_{(7)}  ST $& $ T {\bf{M}}_{(7)}  STS $ & $T^2 {\bf{M}}_{(7)} S T  $ & $ \cdots$ & $ T^6 {\bf{M}}_{(7)}  ST $ & $ S {\bf{M}}_{(7)} T $
 \\ \cline{1-7}
 
 $(1,0,1,1)$ &$ {\bf{M}}_{(7)}  STS  $& $T {\bf{M}}_{(7)}  ST $  & $ T^2 {\bf{M}}_{(7)}  STS$ & $ \cdots$ & $ T^6 {\bf{M}}_{(7)}  STS $ & $ S {\bf{M}}_{(7)}  TS $ \\
 \cline{1-7}
  \end{tabular} 
  \end{center}
  \caption{Classification by the mod 2 structure for $D=7$}
  \label{tb:D=7-mod2}
\end{table}

\subsection{$D=4$}
Here we show the classifications of $\bf{N}$ when $D=4$. We get $7(=1+2+4)$ Types as shown in Table \ref{tab:D=4}. 
We define ${\bf{M}}_{(4)}$ and ${\bf{M}'}_{(4)}$ by
\begin{equation}
{\bf{M}}_{(4)} = 
 \begin{pmatrix}
  4 & 0 \\
  0 & 1
 \end{pmatrix}, \quad 
{\bf{M}}_{(4)}' = 
 \begin{pmatrix}
  2 & 0 \\
  0 & 2
 \end{pmatrix}.
\end{equation}
Type VII is peculiar because one cannot align $\vec{j}$ in a straight line, instead they form a square given by 
\begin{equation}
    \vec{j} = (0,0),\ (0, 1/2),\ (1/2, 0),\ (1/2, 1/2).
\end{equation}
\begin{table}[H]
\begin{center}
\begin{tabular}{|c|c|c|c|} \hline
Type  & ${\bf{N}}$ & gradient \\ \hline
I & $ \left\{ {\bf{M}}_{(4)} \Gamma \right\}$ & 0  \\ \hline
II & $\left\{ T{\bf{M}}_{(4)} \Gamma \right\}$ & $-1$ \\ \hline
III& $\left\{ T^2{\bf{M}}_{(4)} \Gamma \right\}$ & $-2$ \\ \hline
IV& $\left\{ T^3 {\bf{M}}_{(4)} \Gamma \right\}$ & $-3$ \\ \hline
V & $ \left\{ S {\bf{M}}_{(4)} \Gamma \right\}$ & $\infty$  \\ \hline
VI & $\left\{ ST^2{\bf{M}}_{(4)} \Gamma \right\}$ & $1/2$ \\ \hline
VII & $\left\{ {\bf{M}'}_{(4)} \Gamma \right\}$ & $-$ \\ \hline
\end{tabular}
\end{center}
\caption{Types for $D=4$.}
\label{tab:D=4}
\end{table}

Next, let us consider classifications in terms of the mod 2 structure. The most simple one is Type VII because its mod 2 structure is always $(0,0,0,0)$. For other Types, we need a careful treatment. First, we concentrate on the set of $\bf{N}$ which belongs to Type I and have mod 2 structure $(0,0,0,1)$. In general, we can write such ${\bf{N}}$ as follows,
\begin{equation}
{\bf{N}} =
 \begin{pmatrix}
 4 m & 4 n \\ * & * 
 \end{pmatrix},\qquad {\rm gcd}(m, n)=1,\ m \neq 0.
\end{equation}
We act an element of $\Gamma_1(2)$ to the right of ${\bf{N}}$ as
\begin{equation}
%\label{eq: process_a}
 \begin{pmatrix}
 4 m & 4 n \\ * & * 
 \end{pmatrix}
 \begin{pmatrix}
 1 & 0 \\ 2k & 1 
 \end{pmatrix}
 = 
  \begin{pmatrix}
 4 (m+2nk) & 4n \\ * & * 
 \end{pmatrix} =: 
  \begin{pmatrix}
 4 m' & 4n \\ * & * 
 \end{pmatrix}.
\end{equation}
If $n\neq0$, by the appropriate choice of $k \in \mathbb{Z}$, we have
\begin{equation}
|m'| \leq |n|.
\end{equation}
We know $ m' \neq 0$ because of the conservation of mod 2 structure. We consecutively act another element of $\Gamma_1(2)$ to the right as 
\begin{equation}
  \begin{pmatrix}
 4 m' & 4n \\ * & * 
 \end{pmatrix}
   \begin{pmatrix}
 1 & l  \\ 0 & 1 
 \end{pmatrix}
 = 
    \begin{pmatrix}
 4m' & 4 (n + lm')  \\ * & * 
 \end{pmatrix}
 =:
     \begin{pmatrix}
 4m' & 4 n'  \\ * & * 
 \end{pmatrix}.
\end{equation} 
By the appropriate choice of $l \in \mathbb{Z}$, we have
\begin{equation}
|n'| \leq \frac{|m'|}{2}.
\end{equation}
If $n' \neq 0$, we just repeat the process and obtain
\begin{equation}
    |m''| \leq |n'| \leq \frac{|m'|}{2},
\end{equation}
where $m'' \neq 0$ is the new $(1,1)$ element of ${\bf{N} }\Gamma_1(2)$.
After some iterations and the multiplication of $-I \in \Gamma_1(2)$ if necessary, we reach
\begin{equation}
{\bf{N}} \gamma = 
 \begin{pmatrix}
 4  & 0 \\ 2s & 1
 \end{pmatrix},\qquad s \in \mathbb{Z},\  \gamma \in \Gamma_1(2).
\end{equation}
Here we used the fact that actions of $\Gamma_1(2)$ do not change the mod 2 structure. Finally, we act 
\begin{equation}
\begin{pmatrix}
1 & 0 \\ -2s & 1
\end{pmatrix} \in \Gamma_1(2) ,
\end{equation}
from the right, and we get 
\begin{equation}
{\bf{N}} = {\bf{M}}_{(4)} \gamma, \quad \gamma \in \Gamma_1(2).
\end{equation}
Consequently, set of all $\bf{N} \in ({\rm Type\ I})$ with mod 2 structure $(0,0,0,1)$ is given by $\{{\bf{M}}_{(4)} \Gamma_1(2) \}$. It is straightforward  to write similar expressions for other patterns of mod 2 structure. For example, set of $\bf{N} \in ({\rm Type\ I})$ with mod 2 structure $(0,0,1,0)$ is given by 
\begin{equation}
    \{ {\bf{M}}_{(4)} \Gamma_1(2) S \} =  \{ {\bf{M}}_{(4)} S S^{-1} \Gamma_1(2) S \} =  \{ {\bf{M}}_{(4)}S \Gamma^1(2)  \}. 
\end{equation}
Similarly, set of $\bf{N} \in {\rm Type\ I}$ with mod 2 structure $(0,0,1,1)$ is given by 
\begin{equation}
    \{ {\bf{M}}_{(4)} \Gamma_1(2) ST \} =  \{ {\bf{M}}_{(4)} ST (ST)^{-1} \Gamma_1(2) ST \} =  \{ {\bf{M}}_{(4)}ST \Lambda  \}. 
\end{equation}
In order to move to different Types, we just need to act $\Gamma$ to the left of $\bf{N}$.
Results are shown in Tables \ref{tb:D=4-mod2-1}, \ref{tb:D=4-mod2-2}, and \ref{tb:D=4-mod2-3}.

\begin{table}[H]
\begin{center}
  \begin{tabular}{|c||c|c|} \hline 
      \begin{tabular}{c} mod 2 structure\\ $(n_{11}, n_{12}, n_{21}, n_{22})$ \end{tabular} &
      Type I &  
    Type III
   \\ \hline \hline
    $ (0,0,0,1) $& $ \{ {\bf{M}}_{(4)} \Gamma_1(2) \}   $ & $\{ T^2 {\bf{M}}_{(4)} \Gamma_1(2) \} $ 
     \\ \hline   
    $ (0,0,1,0) $ &  $ \{ {\bf{M}}_{(4)} S \Gamma^1(2) \}  $ &  $ \{T^2 {\bf{M}}_{(4)} S \Gamma^1(2) \} $
     \\ \hline    
   $ (0,0,1,1) $ & $ \{{\bf{M}}_{(4)} ST \Lambda \} $   &  $ \{ T^2 {\bf{M}}_{(4)} ST \Lambda \}$ 
    \\ \hline
  \end{tabular} 
  \end{center}
  \caption{Classification of Types I and III for $D=4$}
  \label{tb:D=4-mod2-1}
\end{table}

\begin{table}[H]
\begin{center}
  \begin{tabular}{|c||c|c|} \hline 
      \begin{tabular}{c} mod 2 structure\\ $(n_{11}, n_{12}, n_{21}, n_{22})$ \end{tabular} &
      Type II &  
    Type IV
   \\ \hline \hline
    $ (0,1,0,1) $& $ \{ T {\bf{M}}_{(4)} \Gamma_1(2) \}   $ & $\{ T^3 {\bf{M}}_{(4)} \Gamma_1(2) \} $ 
     \\ \hline   
    $ (1,0,1,0) $ &  $ \{ T {\bf{M}}_{(4)} S \Gamma^1(2) \}  $ &  $ \{T^3 {\bf{M}}_{(4)} S \Gamma^1(2) \} $
     \\ \hline    
   $ (1,1,1,1) $ & $ \{ T {\bf{M}}_{(4)} ST \Lambda \} $   &  $ \{ T^3 {\bf{M}}_{(4)} ST \Lambda \}$ 
    \\ \hline
  \end{tabular} 
  \end{center}
  \caption{Classification of Types II and IV for $D=4$}
  \label{tb:D=4-mod2-2}
\end{table}

\begin{table}[H]
\begin{center}
  \begin{tabular}{|c||c|c|} \hline 
      \begin{tabular}{c} mod 2 structure\\ $(n_{11}, n_{12}, n_{21}, n_{22})$ \end{tabular} &
      Type V &  
    Type VI
   \\ \hline \hline
    $ (0,1,0,0) $& $ \{ S {\bf{M}}_{(4)} \Gamma_1(2) \}   $ & $\{ ST^2 {\bf{M}}_{(4)} \Gamma_1(2) \} $ 
     \\ \hline   
    $ (1,0,0,0) $ &  $ \{ S {\bf{M}}_{(4)} S \Gamma^1(2) \}  $ &  $ \{S T^2 {\bf{M}}_{(4)} S \Gamma^1(2) \} $
     \\ \hline    
   $ (1,1,0,0) $ & $ \{ S {\bf{M}}_{(4)} ST \Lambda \} $   &  $ \{ ST^2 {\bf{M}}_{(4)} ST \Lambda \}$ 
    \\ \hline
  \end{tabular} 
  \end{center}
  \caption{Classifiction of Types V and VI for $D=4$}
  \label{tb:D=4-mod2-3}
\end{table}
One  can  understand  the  reason  why  we  obtain  3  different  patterns  of  mod  2  structure  for each Type (except Type VII) in the following way. Let us focus on Type I as an example. 
The mod 2 structure of the form $(0,0,0,1)$ is invariant only under $\Gamma_1(2) \in \Gamma$. We can classify $\Gamma$ into cosets. The index is given by $[\Gamma : \Gamma_1(2)]=3$ which is the reason of 3 patterns of mod 2 structure \cite{Diamond:2005}. 

\subsection{$D=6$}
Here we show the classifications of $\bf{N}$ when $D=6$. We get $12(=1+2+3+6)$ Types as shown in Table \ref{tab:D=6}.
We define
${\bf{M}}_{(6)}$ by
\begin{equation}
{\bf{M}}_{(6)} = 
 \begin{pmatrix}
  6 & 0 \\
  0 & 1
 \end{pmatrix}.
\end{equation}

\begin{table}[h]
\begin{center}
\begin{tabular}{|c|c|c|c|} \hline
Type  & ${\bf{N}}$ & gradient \\ \hline
I & $ \left\{ {\bf{M}}_{(6)} \Gamma \right\}$ & 0  \\ \hline
II & $\left\{ T{\bf{M}}_{(6)} \Gamma \right\}$ & $-1$ \\ \hline
III& $\left\{ T^2{\bf{M}}_{(6)} \Gamma \right\}$ & $-2$ \\ \hline
IV& $\left\{ T^3 {\bf{M}}_{(6)} \Gamma \right\}$ & $-3$ \\ \hline
V & $ \left\{ T^4 {\bf{M}}_{(6)} \Gamma \right\}$ & $-4$  \\ \hline
VI & $\left\{ T^5{\bf{M}}_{(6)} \Gamma \right\}$ & $-5$ \\ \hline
VII & $\left\{ S {\bf{M}}_{(6)} \Gamma \right\}$ & $\infty$ \\ \hline
VIII & $ \left\{ ST^2 {\bf{M}}_{(6)} \Gamma \right\}$ & $1/2$  \\ \hline
IX & $ \left\{ ST^3 {\bf{M}}_{(6)} \Gamma \right\}$ & $1/3$  \\ \hline
X & $ \left\{ ST^4 {\bf{M}}_{(6)} \Gamma \right\}$ & $1/4$  \\ \hline
XI & $ \left\{ TST^3 {\bf{M}}_{(6)} \Gamma \right\}$ & $-2/3$  \\ \hline
XII & $ \left\{ STST^3 {\bf{M}}_{(6)} \Gamma \right\}$ & $3/2$  \\ \hline
\end{tabular}
\end{center}
\caption{Types for $D=6$.}
\label{tab:D=6}
\end{table}

We further classify in terms of mod 2 structure. 
Results are shown in Tables \ref{tb:D=6-mod-1}, \ref{tb:D=6-mod-2}, \ref{tb:D=6-mod-3}.
\begin{table}[H]
\setlength{\tabcolsep}{1mm} 
\begin{center}
  \begin{tabular}{|c||c|c|c|c|} \hline 
      \begin{tabular}{c} mod 2 structure\\ $(n_{11}, n_{12}, n_{21}, n_{22})$ \end{tabular} &
      Type I & Type III &  Type V & Type XI
   \\ \hline \hline
    $ (0,0,0,1) $& $ \{ {\bf{M}}_{(6)} \Gamma_1(2) \}   $ & $\{ T^2 {\bf{M}}_{(6)} \Gamma_1(2) \} $ & $\{ T^4 {\bf{M}}_{(6)} \Gamma_1(2) \} $ & $\{ T S T^3 {\bf{M}}_{(6)} \Gamma_1(2) \} $
     \\ \hline   
    $ (0,0,1,0) $ &  $ \{ {\bf{M}}_{(6)} S \Gamma^1(2) \}  $ &  $ \{T^2 {\bf{M}}_{(6)} S \Gamma^1(2) \} $ & $ \{T^4 {\bf{M}}_{(6)} S \Gamma^1(2) \} $ & $ \{T S T^3 {\bf{M}}_{(6)} S \Gamma^1(2) \} $
     \\ \hline    
   $ (0,0,1,1) $ & $ \{{\bf{M}}_{(6)} ST \Lambda \} $   &  $ \{ T^2 {\bf{M}}_{(6)} ST \Lambda \}$ &  $ \{ T^4 {\bf{M}}_{(6)} ST \Lambda \}$ &  $\{ T S T^3 {\bf{M}}_{(6)} ST \Lambda \}$
    \\ \hline
  \end{tabular} 
  \end{center}
  \caption{Classification of Types I, III, V, XI for $D=6$}
  \label{tb:D=6-mod-1}
\end{table}

\begin{table}[H]
\setlength{\tabcolsep}{0.8mm} 
\begin{center}
  \begin{tabular}{|c||c|c|c|c|} \hline 
      \begin{tabular}{c} mod 2 structure\\ $(n_{11}, n_{12}, n_{21}, n_{22})$ \end{tabular} &
      Type VII & Type VIII &  Type X & Type XII
   \\ \hline \hline
    $ (0,1,0,0) $& $ \{ S {\bf{M}}_{(6)} \Gamma_1(2) \}   $ & $\{S T^2 {\bf{M}}_{(6)} \Gamma_1(2) \} $ & $\{S T^4 {\bf{M}}_{(6)} \Gamma_1(2) \} $ & $\{S T S T^3 {\bf{M}}_{(6)} \Gamma_1(2) \} $
     \\ \hline   
    $ (1,0,0,0) $ &  $ \{ S{\bf{M}}_{(6)} S \Gamma^1(2) \}  $ &  $ \{S T^2 {\bf{M}}_{(6)} S \Gamma^1(2) \} $ & $ \{ST^4 {\bf{M}}_{(6)} S \Gamma^1(2) \} $ & $ \{ST S T^3 {\bf{M}}_{(6)} S \Gamma^1(2) \} $
     \\ \hline    
   $ (1,1,0,0) $ & $ \{S{\bf{M}}_{(6)} ST \Lambda \} $   &  $ \{ ST^2 {\bf{M}}_{(6)} ST \Lambda \}$ &  $ \{ ST^4 {\bf{M}}_{(6)} ST \Lambda \}$ &  $\{S T S T^3 {\bf{M}}_{(6)} ST \Lambda \}$
    \\ \hline
  \end{tabular} 
  \end{center}
  \caption{Classification of Types VII, VIII, X, XII for $D=6$}
  \label{tb:D=6-mod-2}
\end{table}

\begin{table}[H]
\setlength{\tabcolsep}{1mm} 
\begin{center}
  \begin{tabular}{|c||c|c|c|c|} \hline 
      \begin{tabular}{c} mod 2 structure\\ $(n_{11}, n_{12}, n_{21}, n_{22})$ \end{tabular} &
      Type II & Type IV &  Type VI & Type IX
   \\ \hline \hline
    $ (0,1,0,1) $& $ \{ T {\bf{M}}_{(6)} \Gamma_1(2) \}   $ & $\{T^3 {\bf{M}}_{(6)} \Gamma_1(2) \} $ &  $\{T^5 {\bf{M}}_{(6)} \Gamma_1(2) \} $ &$\{S T^3 {\bf{M}}_{(6)} \Gamma_1(2) \} $
     \\ \hline   
    $ (1,0,1,0) $ &  $ \{ T{\bf{M}}_{(6)} S \Gamma^1(2) \}  $ &  $ \{T^3 {\bf{M}}_{(6)} S \Gamma^1(2) \} $ & $ \{T^5 {\bf{M}}_{(6)} S \Gamma^1(2) \} $ & $ \{ST^3 {\bf{M}}_{(6)} S \Gamma^1(2) \} $
     \\ \hline    
   $ (1,1,1,1) $ & $ \{T {\bf{M}}_{(6)} ST \Lambda \} $   &  $ \{ T^3 {\bf{M}}_{(6)} ST \Lambda \}$ & $\{T^5 {\bf{M}}_{(6)} ST \Lambda \}$  &  $ \{ ST^3 {\bf{M}}_{(6)} ST \Lambda \}$
    \\ \hline
  \end{tabular} 
  \end{center}
  \caption{Classification of Types II, IV, VI, IX  for $D=6$}
  \label{tb:D=6-mod-3}
\end{table}

\subsection{$D=8$}
Here we show the classifications of $\bf{N}$ when $D=8$. We get $15(=1+2+4+8)$ Types as shown in Table \ref{tab:D=8}.
We define ${\bf{M}}_{(8)}$ and ${\bf{M}'}_{(8)}$ by
\begin{equation}
{\bf{M}}_{(8)} = 
 \begin{pmatrix}
  8 & 0 \\
  0 & 1
 \end{pmatrix}, \quad 
{\bf{M}}_{(8)}' = 
 \begin{pmatrix}
  4 & 0 \\
  0 & 2
 \end{pmatrix}.
\end{equation}
\begin{table}[H]
\begin{center}
\begin{tabular}{|c|c|c|c|} \hline
Type  & ${\bf{N}}$ & gradient \\ \hline
I & $ \left\{ {\bf{M}}_{(8)} \Gamma \right\}$ & 0  \\ \hline
II & $\left\{ T{\bf{M}}_{(8)} \Gamma \right\}$ & $-1$ \\ \hline
III& $\left\{ T^2{\bf{M}}_{(8)} \Gamma \right\}$ & $-2$ \\ \hline
IV& $\left\{ T^3 {\bf{M}}_{(8)} \Gamma \right\}$ & $-3$ \\ \hline
V & $ \left\{ T^4 {\bf{M}}_{(8)} \Gamma \right\}$ & $-4$  \\ \hline
VI & $\left\{ T^5{\bf{M}}_{(8)} \Gamma \right\}$ & $-5$ \\ \hline
VII & $ \left\{ T^6 {\bf{M}}_{(8)} \Gamma \right\}$ & $-6$  \\ \hline
VIII & $\left\{ T^7 {\bf{M}}_{(8)} \Gamma \right\}$ & $-7$ \\ \hline
IX & $\left\{ S {\bf{M}}_{(8)} \Gamma \right\}$ & $\infty$ \\ \hline
X & $ \left\{ ST^2 {\bf{M}}_{(8)} \Gamma \right\}$ & $1/2$  \\ \hline
XI & $ \left\{ ST^4 {\bf{M}}_{(8)} \Gamma \right\}$ & $1/4$  \\ \hline
XII & $ \left\{ ST^6 {\bf{M}}_{(8)} \Gamma \right\}$ & $1/6$  \\ \hline
XIII & $ \left\{ {\bf{M}'}_{(8)} \Gamma \right\}$ & $-$  \\ \hline
XIV & $ \left\{ T {\bf{M}'}_{(8)} \Gamma \right\}$ & $-$  \\ \hline
XV & $ \left\{ S {\bf{M}'}_{(8)} \Gamma \right\}$ & $-$  \\ \hline
\end{tabular}
\end{center}
\caption{Types for $D=8$.}
\label{tab:D=8}
\end{table}

We further classify in terms of mod 2 structure. 
Results are shown in Tables \ref{tb:D=8-mod-1}, \ref{tb:D=8-mod-2}, \ref{tb:D=8-mod-3}, \ref{tb:D=8-mod-4}.
\begin{table}[H]
\setlength{\tabcolsep}{1mm} 
\begin{center}
  \begin{tabular}{|c||c|c|c|c|} \hline 
      \begin{tabular}{c} mod 2 structure\\ $(n_{11}, n_{12}, n_{21}, n_{22})$ \end{tabular} &
      Type I & Type III &  Type V & Type VII
   \\ \hline \hline
    $ (0,0,0,1) $& $ \{ {\bf{M}}_{(8)} \Gamma_1(2) \}   $ & $\{ T^2 {\bf{M}}_{(8)} \Gamma_1(2) \} $ & $\{ T^4 {\bf{M}}_{(8)} \Gamma_1(2) \} $ & $\{ T^6 {\bf{M}}_{(8)} \Gamma_1(2) \} $
     \\ \hline   
    $ (0,0,1,0) $ &  $ \{ {\bf{M}}_{(8)} S \Gamma^1(2) \}  $ &  $ \{T^2 {\bf{M}}_{(8)} S \Gamma^1(2) \} $ & $ \{T^4 {\bf{M}}_{(8)} S \Gamma^1(2) \} $ & $ \{T^6 {\bf{M}}_{(8)} S \Gamma^1(2) \} $
     \\ \hline    
   $ (0,0,1,1) $ & $ \{{\bf{M}}_{(8)} ST \Lambda \} $   &  $ \{ T^2 {\bf{M}}_{(8)} ST \Lambda \}$ &  $ \{ T^4 {\bf{M}}_{(8)} ST \Lambda \}$ &  $\{ T^6 {\bf{M}}_{(8)} ST \Lambda \}$
    \\ \hline
  \end{tabular} 
  \end{center}
  \caption{Classification of Types I, III, V, VII for $D=8$}
  \label{tb:D=8-mod-1}
\end{table}

\begin{table}[H]
\setlength{\tabcolsep}{1mm} 
\begin{center}
  \begin{tabular}{|c||c|c|c|c|} \hline 
      \begin{tabular}{c} mod 2 structure\\ $(n_{11}, n_{12}, n_{21}, n_{22})$ \end{tabular} &
      Type IX & Type X &  Type XI & Type XII
   \\ \hline \hline
    $ (0,1,0,0) $& $ \{ S {\bf{M}}_{(8)} \Gamma_1(2) \}   $ & $\{S T^2 {\bf{M}}_{(8)} \Gamma_1(2) \} $ & $\{S T^4 {\bf{M}}_{(8)} \Gamma_1(2) \} $ & $\{S T^6 {\bf{M}}_{(8)} \Gamma_1(2) \} $
     \\ \hline   
    $ (1,0,0,0) $ &  $ \{ S{\bf{M}}_{(8)} S \Gamma^1(2) \}  $ &  $ \{S T^2 {\bf{M}}_{(8)} S \Gamma^1(2) \} $ & $ \{ST^4 {\bf{M}}_{(8)} S \Gamma^1(2) \} $ & $ \{S T^6 {\bf{M}}_{(8)} S \Gamma^1(2) \} $
     \\ \hline    
   $ (1,1,0,0) $ & $ \{S{\bf{M}}_{(8)} ST \Lambda \} $   &  $ \{ ST^2 {\bf{M}}_{(8)} ST \Lambda \}$ &  $ \{ ST^4 {\bf{M}}_{(8)} ST \Lambda \}$ &  $\{S T^6 {\bf{M}}_{(8)} ST \Lambda \}$
    \\ \hline
  \end{tabular} 
  \end{center}
  \caption{Classification of Types IX, X, XI, XII for $D=8$}
  \label{tb:D=8-mod-2}
\end{table}

\begin{table}[H]
\setlength{\tabcolsep}{1mm} 
\begin{center}
  \begin{tabular}{|c||c|c|c|c|} \hline 
      \begin{tabular}{c} mod 2 structure\\ $(n_{11}, n_{12}, n_{21}, n_{22})$ \end{tabular} &
      Type II & Type IV &  Type VI & Type VIII
   \\ \hline \hline
    $ (0,1,0,1) $& $ \{ T {\bf{M}}_{(8)} \Gamma_1(2) \}   $ & $\{T^3 {\bf{M}}_{(8)} \Gamma_1(2) \} $ & $\{T^5 {\bf{M}}_{(8)} \Gamma_1(2) \} $ & $\{T^7 {\bf{M}}_{(8)} \Gamma_1(2) \} $
     \\ \hline   
    $ (1,0,1,0) $ &  $ \{ T{\bf{M}}_{(8)} S \Gamma^1(2) \}  $ &  $ \{T^3 {\bf{M}}_{(8)} S \Gamma^1(2) \} $ & $ \{T^5 {\bf{M}}_{(8)} S \Gamma^1(2) \} $ & $ \{T^7 {\bf{M}}_{(8)} S \Gamma^1(2) \} $
     \\ \hline    
   $ (1,1,1,1) $ & $ \{T {\bf{M}}_{(8)} ST \Lambda \} $   &  $ \{ T^3 {\bf{M}}_{(8)} ST \Lambda \}$ &  $ \{ T^5 {\bf{M}}_{(8)} ST \Lambda \}$ &  $\{T^7 {\bf{M}}_{(8)} ST \Lambda \}$
    \\ \hline
  \end{tabular} 
  \end{center}
  \caption{Classification of Types II, IV, VI, VIII for $D=8$}
  \label{tb:D=8-mod-3}
\end{table}

\begin{table}[H]
\begin{center}
  \begin{tabular}{|c||c|c|c|} \hline 
      \begin{tabular}{c} mod 2 structure\\ $(n_{11}, n_{12}, n_{21}, n_{22})$ \end{tabular} &
      Type XIII & Type XIV &  Type XV 
   \\ \hline \hline
    $ (0,0,0,0) $& $ \{ {\bf{M}}_{(8)}' \Gamma \}   $ & $\{ T{\bf{M}}_{(8)}' \Gamma \} $ & $\{ S {\bf{M}}_{(8)}' \Gamma \} $      \\ \hline   
  \end{tabular} 
  \end{center}
  \caption{Classification of Types XIII, XIV, XV for $D=8$}
  \label{tb:D=8-mod-4}
\end{table}

One can understand the reason why we obtain 3 different Types corresponding to the mod 2 structure $(0,0,0,0)$ as shown in Table \ref{tb:D=8-mod-4}. This is because the situation is just equivalent to the classifications of $\bf{N}$ with $D=2$. 

\section{Higgs sector}
Here we study the Higgs bosons which can couple to fermions under the $T^4/Z_2$ compactification models. We require that the generation numbers for both left- and right-handed fermions are three as in the SM. For this purpose, let us consider Yukawa couplings among left- and right-handed fermions with Higgs boson,
\begin{equation}
\label{eq: YukawaT4Z2}
Y_{\vec{i}, \vec{j}, \vec{k}} \propto \int_{T^4/Z_2} d^2z d^2 \bar{z}\  \psi_{T^4/Z_2, \pm}^{(\vec{i}+\vec{\alpha}_L {\bf{N}}_L^{-1}, \vec{\beta}_L)} (\vec{z}, \Omega) \psi_{T^4/Z_2, \pm}^{(\vec{j}+\vec{\alpha}_R {\bf{N}}_R^{-1}, \vec{\beta}_R)} (\vec{z}, \Omega) \left( \psi_{T^4/Z_2, \pm}^{(\vec{k}+\vec{\alpha}_H {\bf{N}}_H^{-1}, \vec{\beta}_H)} (\vec{z}, \Omega) \right)^*.
\end{equation}
Here we concentrate on the case when all wave functions have positive chirality as in Eq.$(\ref{eq: zero-mode})$. 
From the gauge invariance we need
\begin{equation}
    {\bf{N}}_H = {\bf{N}}_L + {\bf{N}}_R .
\end{equation}
Moreover, note that the following conditions need to be satisfied to obtain non-vanishing Yukawa couplings,
\begin{align}
\label{eq: SS1_Yukawa}
\vec{\alpha}_L + \vec{\alpha}_R &\equiv \vec{\alpha}_H\quad {(\rm mod 1)}, \\
\label{eq: SS2_Yukawa}
\vec{\beta}_L + \vec{\beta}_R &\equiv \vec{\beta}_H\quad {(\rm mod 1)}, \\
\label{eq: parity_Yukawa}
p_L p_R &= p_{H},
\end{align}
where $p_L, p_R,$ and $p_H$ denote the $Z_2$ parity($\pm 1$) of left, right, and Higgs sector wave functions. For example, these conditions can be understood in the following way. Consider $\vec{z} + \vec{e}_n$, then we find
\begin{equation}
Y_{\vec{i}, \vec{j}, \vec{k}}  = e^{2\pi i (\alpha^n_L + \alpha^n_R-\alpha^n_H)}Y_{\vec{i}, \vec{j}, \vec{k}},
\end{equation}
from which we obtain Eq.$(\ref{eq: SS1_Yukawa})$. Similarly, taking $\vec{z}$ to $\vec{z} + \Omega \vec{e}_n$ or $- \vec{z}$ provides the conditions Eqs.$(\ref{eq: SS2_Yukawa})$ and ($\ref{eq: parity_Yukawa}$).
 
It is a good starting point to suppose that the zero-modes of the left-handed fermions in $T^4/Z_2$ are under the magnetic flux given by
\begin{equation}
\label{eq: Nl}
{\bf{N}}_L = 
 \begin{pmatrix}
 n_1 & n_2 \\ n_3 & n_4
 \end{pmatrix}
 = \begin{pmatrix}
 D & k \\
 0 & 1
 \end{pmatrix} \gamma, \quad \gamma \in \Gamma ,
\end{equation}
where $k=\{0,1,..,D-1 \}$. Then we find 
\begin{align}
{\bf{N}}_L \Omega = 
 \begin{pmatrix}
 D & k \\
 0 & 1
 \end{pmatrix} \gamma
  \begin{pmatrix}
 \omega_1 & \omega_2 \\
 \omega_3 & \omega_4
 \end{pmatrix}
 =  \begin{pmatrix}
 D & k \\
 0 & 1
 \end{pmatrix}
 \begin{pmatrix}
 \omega_1' & \omega_2' \\
 \omega_3' & \omega_4'
 \end{pmatrix}
 = 
 \begin{pmatrix}
 D \omega_1' + k \omega_3' & D \omega_2' + k \omega_4' \\
 \omega_3' & \omega_4'
 \end{pmatrix}.
\end{align}
In order to satisfy the SUSY condition, we need 
\begin{equation}
\label{eq: L_SUSY}
\omega_3' = D \omega_2' + k \omega_4' .
\end{equation}
On the other hand, we denote the magnetic flux in Dirac equation of the right-handed fermions as 
\begin{equation}
{\bf{N}}_R = 
 \begin{pmatrix}
 m_1 & m_2 \\ m_3 & m_4
 \end{pmatrix}.
\end{equation}
Then we get
\begin{align}
 \begin{aligned}
{\bf{N}}_R \Omega = {\bf{N}}_R \gamma^{-1} \Omega' 
=
\begin{pmatrix}
m_1' & m_2' \\
m_3' & m_4'
\end{pmatrix}
 \begin{pmatrix}
 \omega_1' & \omega_2' \\
 \omega_3' & \omega_4'
 \end{pmatrix}.
 \end{aligned}
\end{align}
In order to satisfy the SUSY condition, we need
\begin{equation}
\label{eq: R_SUSY}
m_1' \omega_2' + m_2' \omega_4' = m_3' \omega_1' + m_4' \omega_3'.
\end{equation}
From Eqs.$(\ref{eq: L_SUSY})$ and $(\ref{eq: R_SUSY})$, we obtain
\begin{equation}
m_3' = \frac{(m_1' - D m_4') \omega_2' + (m_2' - k m_4') \omega_4'}{\omega_1'}.
\end{equation}
Note that 
$m_i',\ (i=,1,2,3,4)$ are integers whereas $\omega_i'$ are complex numbers. It is then natural to demand
\begin{equation}
\label{eq: SUSY_sufficient}
m_1' = Dm_4', \quad m_2' = km_4'.
\end{equation} 
Otherwise the complex structure matrix $\Omega$ must satisfy some very special relations such as $\omega_i' / \omega_1' \in \mathbb{Z},\ (i=2,3,4)$ which strongly constrain the continuous degrees of freedom in $\Omega$. Here we analyze the Higgs sector under the conditions Eq.($\ref{eq: SUSY_sufficient}$). Then we get  

\begin{equation}
{\bf{N}}_R \gamma^{-1} = m_4'
 \begin{pmatrix}
 D & k \\
 0 & 1
 \end{pmatrix} = m_4' {\bf{N}}_L \gamma^{-1},
\end{equation}
showing that ${\bf{N}}_R$ is simply given by $m_4' {\bf{N}}_L$. It is obvious from Table $\ref{tb:alpha0D}$ and $\ref{tb:alpha1D}$ that $m_4'$ needs to be 1, i.e.,
\begin{equation}
\label{eq: L=R}
    {\bf{N}}_L = {\bf{N}}_R .
\end{equation}
It seems clear that we get the same conclusion even if ${\bf{N}}_L$ is not given by Eq.$(\ref{eq: Nl})$. 
As a result, ${\bf{N}}_H$ is obtained as 
\begin{equation}
{\bf{N}}_H = 2{\bf{N}}_L. 
\end{equation}
 Now, we have all the information necessary to determine the Higgs sector.

Let us first look at the case when $D= {\rm det{\bf{N}}}_L = 4 (= {\rm det} {\bf{N}}_R)$. Assignments of quantum numbers yielding 3 degenerated zero-modes of fermions are summarized in Tables \ref{tb:HiggsD4} and \ref{tab:mod-2}. 
\begin{table}[H]
    \begin{minipage}[t]{.45\textwidth}
    \begin{center}
    \begin{tabular}{c|c|c|c}
        $p_L$ & $\vec{\alpha}_L$ & $\vec{\beta}_L$ & mod 2 structure \\ \hline
        \multirow{7}{*}{$1$}  & \multirow{4}{*}{$(0,0)$} & $(0,0)$ & \maru{2} $-$ \maru{10} \\  \cline{3-4}
            &               &   $(0, \frac{1}{2})$ & \maru{2} $-$ \maru{4}  \\ \cline{3-4}
            &               &   $(\frac{1}{2}, 0)$ & \maru{5} $-$ \maru{7}  \\ \cline{3-4}
            &               &   $(\frac{1}{2},\frac{1}{2})$ & \maru{8} $-$ \maru{10}  \\ \cline{2-4}
            & $(0, \frac{1}{2})$ & $(0,0)$         & \maru{2},\maru{5},\maru{8} \\ \cline{2-4}
            & $(\frac{1}{2}, 0)$ & $(0,0)$         & \maru{3},\maru{6},\maru{9} \\ \cline{2-4}
            & $(\frac{1}{2},\frac{1}{2})$ & $(0,0)$         & \maru{4},\maru{7},\maru{10} \\ \hline
        \multirow{9}{*}{$-1$} & \multirow{3}{*}{$(0, \frac{1}{2})$} & $(0,\frac{1}{2})$ & \maru{2} \\ \cline{3-4} 
            &               & $(\frac{1}{2},0)$   & \maru{5} \\ \cline{3-4}
            &               & $(\frac{1}{2},\frac{1}{2})$   & \maru{8} \\ \cline{2-4}
            & \multirow{3}{*}{$(\frac{1}{2},0)$} & $(0, \frac{1}{2})$ & \maru{3} \\ \cline{3-4} 
            &               & $(\frac{1}{2}, 0)$   & \maru{6} \\ \cline{3-4}
            &               & $(\frac{1}{2},\frac{1}{2})$   & \maru{9} \\ \cline{2-4}      
            & \multirow{3}{*}{$(\frac{1}{2},\frac{1}{2})$} & $(0, \frac{1}{2})$ & \maru{4} \\ \cline{3-4}
            &               & $(\frac{1}{2}, 0)$   & \maru{7} \\ \cline{3-4}
            &               & $(\frac{1}{2},\frac{1}{2})$   & \maru{10} \\ \hline        
    \end{tabular}
    \end{center}
    \caption{3 generation of fermion when $D=4$.
The mod 2 structures are shown in Table \ref{tab:mod-2}.}
    \label{tb:HiggsD4}
 \end{minipage}
 \hfill
 \begin{minipage}[t]{.45\textwidth}
 \begin{center}
     \begin{tabular}{c|c}
                  & $(n_{11}, n_{12}, n_{21}, n_{22})$ \\ \hline
        \maru{1}  &  $(0,0,0,0)$  \\ \hline
        \maru{2}  &  $(0,0,0,1)$ \\ \hline
        \maru{3}  &  $(0,0,1,0)$\\ \hline
        \maru{4}  &  $(0,0,1,1)$\\ \hline
        \maru{5}  &  $(0,1,0,0)$  \\ \hline
        \maru{6}  &  $(1,0,0,0)$  \\ \hline
        \maru{7}  &  $(1,1,0,0)$  \\ \hline
        \maru{8}  &  $(0,1,0,1)$  \\ \hline
        \maru{9}  &  $(1,0,1,0)$  \\ \hline
        \maru{10} &  $(1,1,1,1)$  \\ \hline
        \maru{11} &  $(0,1,1,0)$  \\ \hline
        \maru{12} &  $(1,0,0,1)$  \\ \hline
        \maru{13} &  $(0,1,1,1)$  \\ \hline
        \maru{14} &  $(1,0,1,1)$  \\ \hline
        \maru{15} &  $(1,1,0,1)$  \\ \hline
        \maru{16} &  $(1,1,1,0)$  \\ 
     \end{tabular}
 \end{center}
     \caption{mod 2 structure}
\label{tab:mod-2}
 \end{minipage}
\end{table}
Only in this section, let us count the number of models in the following way. When following four conditions for each three sector(L, R, H) are specified, we count it as one model although there exist infinite number of different ${\bf{N}}$ in it. The four conditions are: $Z_2$ parity $p$, Scherk-Schwarz phases $\vec{\alpha}$, $\vec{\beta}$ and the mod 2 structure. From Table $\ref{tb:HiggsD4}$, we notice that there are 36 ways of assigning the quantum numbers in the left-handed sector. From Eq.$(\ref{eq: L=R})$, we notice that ${\bf {N}}_R$ must have the same mod 2 structure as ${\bf{N}}_L$. Therefore, the number of models is much less than $36 \times 36=1296$, and we only have $144$ models. Once the four conditions for left- and right-handed fermion are specified, we automatically obtain the conditions for the Higgs sector. It is important to note that the mod 2 structure of ${\bf{N}}_H$ is always $(0,0,0,0)$. This makes it easy to read the number of Higgs modes from Tables $\ref{tb:alpha0}$, $\ref{tb:alpha1}$, and Eq.(\ref{eq: generation_counting}). We notice that $C_+ - C_-$
is nonzero only when its Scherk-Schwarz phases are vanishing. Having noted this point, Table $\ref{tb:HiggsD4}$ suggests that both left-and right-handed fermions must have the identical quantum number assignments in order to produce non-vanishing $C_+ - C_-$ in the Higgs sector.  
Therefore, we have only 36 models yielding non-vanishing $C_+ - C_- (=4)$. This explains why we get 36 models yielding 10  Higgs modes as shown in Table $\ref{tb: Higgs_gen_D4}$. All other models correspond to the cases $C_+-C_- =0$, yielding 8 Higgs modes. It is not difficult to extend this kind of analysis for other values of $D$. Results are summarized in Tables, \ref{tab:possible-n3}, \ref{tab:possible-n5}, \ref{tab:possible-n4}, \ref{tab:possible-n2}.
\begin{table}[H]
\begin{center}
\begin{tabular}{|c|c|c|c|} \hline
number of Higgs modes & 8 & 9 & 10  \\ \hline
combination number of $\psi_L$ and $\psi_R$ & 108 & 0 & 36  \\ \hline
\end{tabular}
\end{center}
\caption{$D_L=D_R=4, D_H=16$}
\label{tb: Higgs_gen_D4}
\end{table}

\begin{table}[H]
\begin{center}
\begin{tabular}{|c|c|c|c|} \hline
number of Higgs modes& 10 & 11 & 12  \\ \hline
combination number of $\psi_L$ and $\psi_R$ & 1440 & 0 & 96  \\ \hline
\end{tabular}
\end{center}
\caption{$D_L=D_R=5, D_H=20$}
\label{tab:possible-n3}
\end{table}

\begin{table}[H]
\begin{center}
\begin{tabular}{|c|c|c|c|c|c|} \hline
number of Higgs modes& 10 & 11 & 12 & 13 & 14  \\ \hline
combination number of $\psi_L$ and $\psi_R$ & 328 & 0 & 4048 & 0 & 328  \\ \hline
\end{tabular}
\end{center}
\caption{$D_L=D_R=6, D_H=24$}
\label{tab:possible-n5}
\end{table}

\begin{table}[H]
\begin{center}
\begin{tabular}{|c|c|c|c|} \hline
number of Higgs modes& 14 & 15 & 16  \\ \hline
combination number of $\psi_L$ and $\psi_R$ & 1440 & 0 & 96  \\ \hline
\end{tabular}
\end{center}
\caption{$D_L=D_R=7, D_H=28$}
\label{tab:possible-n4}
\end{table}

\begin{table}[H]
\begin{center}
\begin{tabular}{|c|c|c|c|} \hline
number of Higgs modes& 16 & 17 & 18  \\ \hline
combination number of $\psi_L$ and $\psi_R$ & 108 & 0 & 36  \\ \hline
\end{tabular}
\end{center}
\caption{$D_L=D_R=8, D_H=32$}
\label{tab:possible-n2}
\end{table}

Based on this simple analysis, we found that the number of Higgs modes can be 8, 10, 12, 14, 16, or 18. We can observe similarities between $D_L=4$ and $D_L=8$ cases, as well as between $D_L=5$ and $D_L=7$ cases. This is traced back to Tables $\ref{tb:alpha0D}$ and $\ref{tb:alpha1D}$ showing both $D_+$ and $D_-$ values realizing 3 degeneracy in the fermion sector. We can clearly observe the correspondence. 

At last, we give the explicit form of Yukawa coupling constant in Eq.(\ref{eq: YukawaT4Z2}) in terms of Riemann theta-functions. Yukawa coupling constants in $T^4/Z_2$ models are given by the linear combinations of those computed in $T^4$ models. Yukawa coupling constants in $T^4$ models are written as

\begin{align}
\label{eq: overlap}
Y_{\vec{i}, \vec{j}, \vec{k}}^{(T^4)} \propto \int_{T^4} d^2z d^2 \bar{z}\  \psi^{(\vec{i}+\vec{\alpha}_L {\bf{N}}_L^{-1}, \vec{\beta}_L)} (\vec{z}, \Omega) \psi^{(\vec{j}+\vec{\alpha}_R {\bf{N}}_R^{-1}, \vec{\beta}_R)} (\vec{z}, \Omega) \left( \psi^{(\vec{k}+\vec{\alpha}_H {\bf{N}}_H^{-1}, \vec{\beta}_H)} (\vec{z}, \Omega) \right)^*.
\end{align}
Theta identities proven in Ref.\cite{Antoniadis:2009bg} allow us to evaluate explicitely.\footnote{In Appendix A, we show an explicit derivation.} In general, we can write
\begin{align}
\begin{aligned}
\label{eq: Yukawa_general}
Y_{\vec{i}, \vec{j}, \vec{k}}^{(T^4)}& \sim \sum_{\vec{m}} \vartheta
 \begin{bmatrix}
 ( \vec{i} + \vec{\alpha}_L{\bf{N}}_L^{-1} - \vec{j}  - \vec{\alpha}_R {\bf{N}}_R^{-1} + \vec{m} ) \cdot \frac{{\bf{N}}_L {\bf{N}}_H^{-1} {\bf{N}}_R P^{-1}}{{\rm lcm}(D_L, D_R)} \\
 0
 \end{bmatrix} \\
 &\qquad \left( {\rm lcm}(D_L, D_R) P (- {\bf{N}}_L^{-1} \vec{\beta}_L + {\bf{N}}_R^{-1} \vec{\beta}_R), {\rm lcm}(D_L, D_R)^2 P {\bf{N}}_L^{-1} {\bf{N}}_H {\bf{N}}_R^{-1} \Omega^T P^T \right)  \\
 & \qquad \times \delta_{((\vec{i} + \vec{\alpha}_L {\bf N}^{-1}_L ){\bf{N}}_L + (\vec{j} + \vec{\alpha}_R {\bf N}^{-1}_R) {\bf{N}}_R + \vec{m} {\bf{N}}_L ) {\bf{N}}_H^{-1} ,\vec{k}+\vec{\alpha}_H {\bf{N}}_H^{-1} + \vec{l}}\ ,
 \end{aligned}
\end{align}
where $\vec{l}$ in the Kronecker's delta denotes possible two dimensional integer vectors. This means we take ${\rm mod} ( \vec{e}_n {\bf{N}}_H)$. This kind of selection rule is also present in $T^2$ cases \cite{Cremades:2004wa, Abe:2015yva, Fujimoto:2016zjs}. The least common multiple of $D_L$ and $D_R$ is denoted by ${\rm lcm}(D_L, D_R)$. Note that $P$ is an element of $GL(2, \mathbb{Z})$, and we can choose it freely. The sum of $\vec{m}$ is taken over integer points in the cell spanned by    
\begin{equation}
\label{eq: m_sum}
{\vec{e}}_n^{\ '} = \vec{e}_n {\rm lcm}(D_L, D_R) \cdot P {\bf{N}}_L^{-1} {\bf{N}}_H  {\bf{N}}_R^{-1} ,\quad (n= 1,2).
\end{equation}
We can simplify the expression if ${\bf{N}}_L = {\bf{N}}_R$ as
\begin{align}
\begin{aligned}
Y_{\vec{i}, \vec{j}, \vec{k}}^{(T^4)}& \sim \sum_{\vec{m}} \vartheta
 \begin{bmatrix}
 (\vec{i}  - \vec{j} + \vec{m} + (\vec{\alpha}_L - \vec{\alpha}_R) {\bf{N}}_L^{-1} ) \cdot \frac{{\bf{N}}_L  P^{-1}}{{2 D_L} }\\
 0
 \end{bmatrix} \\
 &\qquad \left( D_L P {\bf{N}}_L^{-1}(-  \vec{\beta}_L +  \vec{\beta}_R), 2 D_L^2 P {\bf{N}}_L^{-1}  \Omega^T P^T \right)  \\
 & \qquad \times \delta_{\vec{i} + \vec{j} + \vec{m} + (\vec{\alpha}_L + \vec{\alpha}_R) {\bf{N}}_L^{-1}, 2\vec{k} + \vec{\alpha}_H {\bf{N}}_L^{-1} + 2 \vec{l}}\ ,
 \end{aligned}
\end{align}
where $\vec{m}$ is summed over integer points in the cell spanned by    
\begin{equation}
{\vec{e}}_n^{\ '} = \vec{e}_n (2 D_L) \cdot P {\bf{N}}_L^{-1} ,\quad (n= 1,2). 
\end{equation}

\section{Yukawa couplings and Types}
Here, we discuss how the classification we carried out is reflected in physical quantities obtained through the magnetized $T^4$ models. 
When $D_L = D_R$ and Scherk-Schwarz phases are vanishing, we can write the Yukawa coupling constant as
\begin{align}
\begin{aligned}
\label{eq: Yukawa_Types}
    Y_{\vec{i}, \vec{j}, \vec{k}}^{(T^4)}& \sim \sum_{\vec{m}} \vartheta
 \begin{bmatrix}
 ( \vec{i} - \vec{j}   + \vec{m} ) \cdot \frac{{\bf N}_H P^{-1}}{{\rm det}{\bf N}_H} \\
 0
 \end{bmatrix} 
 \  \left( 0, {\rm lcm}(D_L, D_R) P ({\rm det}{\bf N}_H) {\bf N}_{H}^{-1} \Omega^T P^T \right)  \\
 & \qquad \times \delta_{\vec{i}{\bf{N}}_L + \vec{j} {\bf{N}}_R - \vec{k}{\bf{N}}_H   , \vec{l}{\bf{N}}_H- \vec{m} {\bf{N}}_L},
\end{aligned}
\end{align}
where $\vec{m}$ are $\mathbb{Z}^2$ vectors inside the cell spanned by
\begin{equation}
\label{eq: m_def}
    {\vec{e}_n}^{\ '} = \vec{e}_n P ({\rm det}{\bf N}_H) {\bf N}_H^{-1}.
\end{equation}
Eqs.(\ref{eq: Yukawa_Types}) and (\ref{eq: m_def}) can be derived from the general form Eqs.(\ref{eq: Yukawa_general})
and (\ref{eq: m_sum}) by use of the following relation 
\begin{equation}
\label{eq: useful}
    D_L {\bf N}_R^{-1} {\bf N}_H {\bf N}_L^{-1} = ({\rm det}{\bf N}_H){\bf N}_H^{-1},
\end{equation}
which is satisfied when $D_L=D_R$.
The flavor structure is governed by the characteristic of the Riemann theta-function and the selection rule.  

Firstly, let us look at the characteristic written as
\begin{equation}
     ( \vec{i} - \vec{j}   + \vec{m} ) \cdot \frac{{\bf N}_H P^{-1}}{{\rm det}{\bf N}_H}.
\end{equation}
This shows the significance of Types of ${\bf N}_L$ and ${\bf N}_R$ since $\vec{i}$ and $\vec{j}$ are present. Next, notice that by a suitable choice of $P \in { GL}(2,\mathbb{Z})$, we can always bring ${\bf N}_H P^{-1}$ into the representative form of our Type. This allows us to understand the flavor structure in a simple way. One may question $\vec{m}$ will be affected by the choice of $P$ and our discussion is ruined. However, $\vec{m}$ can be chosen independent of $P$. For its proof, readers should refer to Appendix \ref{m_and_P}.
Based on this fact, we can further simplify the alignment of $\vec{m}$ by knowing the Type of ${\bf{N}}_H$.
 We discussed about this in Appendix \ref{m_and_Type}.
 
Secondly, let us look at the selection rule,
\begin{equation}
\label{eq: selection}
    \vec{i}{\bf{N}}_L + \vec{j} {\bf{N}}_R - \vec{k} {\bf{N}}_H = \vec{l} {\bf{N}}_H - \vec{m} {\bf{N}}_L.
\end{equation}
 In addition to our Type defined in Eq.(\ref{eq: Type}), it is worth defining L-Type concerning the term $\vec{l} {\bf N}_H$. If ${\bf N}_1$ and ${\bf{N}}_2$ are related by
\begin{equation}
\label{eq: L-Type}
    \gamma {\bf N}_1 = {\bf N}_2,\quad \gamma \in \Gamma=SL(2,\mathbb{Z}),
\end{equation}
we call ${\bf N}_1$ and ${\bf N}_2$ are in the same L-Type. 
Intersection matrices in the same L-Type produce the same ${\rm mod}(\vec{e}_n {\bf{N}})$ structure. This is because $\gamma$ in Eq.(\ref{eq: L-Type}) can be interpreted as the change of lattice vectors which does not alter the lattice points. The number of different L-Types is equal to that of Types for given ${\rm det}{\bf N}_H$ because L-Type can be understood as the transpose of Type. Having identified both Type and L-Type of ${\bf N}_H$, we are able to know the number of solutions satisfying $\vec{m} {\bf N}_L = \vec{l}{\bf N}_H$. This corresponds to the number of terms which survive the selection rule and therefore contribute to $  Y_{\vec{i}, \vec{j}, \vec{k}}^{(T^4)}$. 

Above analysis implies that our Types and L-Types will be important to understand the pattern of the Yukawa couplings.

\section{Conclusion}
\label{conclusion}

We have studied $T^4$ and $T^4/Z_2$ compactifications with magnetic fluxes. We found that Scherk-Schwarz phases can take 16 different values on $T^4/Z_2$. We wrote down zero-mode wave functions including these phases. In  $T^4$ models, the degeneracy of zero-modes is equal to the determinant of the intersection matrix $\bf{N}$. $\bf{N}$ is a $2 \times 2$ integer matrix which determines the orientations and strength of the background magnetic field. In  $T^4/Z_2$ models, the degeneracy is determined by the following four information: $|{\rm det} {\bf{N}}|$, Scherk-Schwarz phases, $Z_2$ parity, and the mod 2 structure of ${\bf{N}}$. The mod 2 structure refers to the mod 2 value of each matrix element. There are a number of three-generation models in both  $T^4$ and  $T^4/Z_2$ cases. Thus, we proposed a method to classify them. We used some group theoretical languages to do it. This method offers a unified understanding over a number of models including physical observables such as the Yukawa coupling constants. Moreover, we studied the number of Higgs modes under the requirement of the SUSY condition. 
Models in each Type have the same aliment of the vectors $\vec j$ on $T^4$.
That leads to the same flavor symmetry, which was studied in Ref.~\cite{Abe:2014nla}.
Similarly, models in each Type have the same number of zero-modes on $T^4/Z_2$ when their mod 2 structure are the same.
In this sense, models in one Type are equivalent to each other when we see only the left-handed fermion sector or 
the right-handed fermion sector.
However, phenomenological aspects, e.g. their Yukawa couplings, depend on models even if each of the left-handed sector and the right handed sector belongs to the same Type. We need information such as the flux of the Higgs sector in order to discuss the phenomenological aspects as we saw in section 6.

The obtained results may be a starting point to extend our analysis to more phenomenological studies in the future.   
One of interesting applications is to study the flavor structure in $T^4/Z_2$ orbifold models as well as $T^4$ models. 
It is intriguing to study the realization of quark and lepton masses and their mixing angles similar to 
analyses on $T^2/Z_2$ orbifold models \cite{Abe:2012fj,Abe:2014vza,Fujimoto:2016zjs,Kobayashi:2016qag,Kikuchi:2021yog}.
Also, it is important to study the flavor symmetry of $T^4/Z_2$ orbifold models.
The modular symmetry of $T^4$ and $T^4/Z_2$ is the symplectic modular symmetry $Sp(4,\mathbb{Z})$, 
which is larger than the modular symmetry of $T^2$ and $T^2/Z_2$~\footnote{Calabi-Yau compactifications have many moduli, and they 
also have symplectic modular symmetries \cite{Strominger:1990pd,Candelas:1990pi,Ishiguro:2020nuf,Ishiguro:2021ccl}.}.
We would study behaviors of zero-mode wave functions under $Sp(4,\mathbb{Z})$ elsewhere.

%%%%%%%%%%%%%%%%%%%%%%%%%%%%%%%%%%%%%%%%%%%%%%%%%%%%
%-------- acknowledgement -------%
\vspace{1.5 cm}
\noindent
{\large\bf Acknowledgement}\\

This work was supported by JST SPRING Grant Number JPMJSP2119(SK) and 
JSPS KAKENHI Grant Number JP20J20388(HU).

%-------- Appendix -------%

\appendix
\section{Computing Yukawa couplings}
Here we give a proof of Eqs.(\ref{eq: Yukawa_general}) and (\ref{eq: m_sum}) based on Ref. \cite{Antoniadis:2009bg}.
\subsection{Theta identity}
When computing Yukawa couplings in magnetized $T^4$ models, it is convenient to use the following theta identity given by
\begin{align}
\begin{aligned}
\label{eq: theta_id}
    &\vartheta 
     \begin{bmatrix}
         \vec{j}_1 \\ 0
     \end{bmatrix}(\vec{z}_1, {\bf{N}}_1 \Omega) 
 \cdot
    \vartheta 
     \begin{bmatrix}
         \vec{j}_2 \\ 0
     \end{bmatrix}(\vec{z}_2, {\bf{N}}_2 \Omega) \\
&= \sum_{\vec{m}} 
    \vartheta 
     \begin{bmatrix}
     (\vec{j}_1 {\bf{N}}_1 + \vec{j}_2 {\bf{N}}_2 + \vec{m} {\bf{N}}_1) ({\bf{N}}_1 + {\bf{N}}_2)^{-1} \\
     0 
     \end{bmatrix}(\vec{z}_1 + \vec{z}_2 , ({\bf{N}}_1+{\bf{N}}_2)\Omega) \\
& \times  
     \vartheta 
     \begin{bmatrix}
     (\vec{j}_1 - \vec{j}_2 + \vec{m}) 
     \frac{{\bf{N}}_1 ({\bf{N}}_1 + {\bf{N}}_2)^{-1} {\bf{N}}_2}{{\rm lcm}(D_1, D_2)}P^{-1} \\
     0 
     \end{bmatrix} \\
&
({\rm lcm}(D_1, D_2) P ({\bf{N}}_1^{-1} \vec{z}_1 - {\bf{N}}_2^{-1} \vec{z}_2) , ({\rm lcm}(D_1, D_2))^2 P {\bf{N}}_1^{-1} ({\bf{N}}_1 + {\bf{N}}_2) {\bf{N}}_2^{-1} \Omega^T P^T),
\end{aligned}
\end{align}
where $P \in GL(2, \mathbb{Z})$.
 The sum of $\vec{m}$ is taken over integer points in the cell spanned by    
\begin{equation}
\label{eq: cell_m}
{\vec{e}}_n^{\ '} = \vec{e}_n {\rm lcm}(D_1, D_2) \cdot P {\bf{N}}_1^{-1} ({\bf{N}}_1 + {\bf{N}}_2)  {\bf{N}}_2^{-1} ,\quad (n= 1,2).
\end{equation}

\paragraph{proof}\ %a% 
\\
The left-hand side of Eq.(\ref{eq: theta_id}) is
\begin{align}
\label{eq: theta_product}
     &\vartheta 
     \begin{bmatrix}
         \vec{j}_1 \\ 0
     \end{bmatrix}(\vec{z}_1, {\bf{N}}_1 \Omega) 
 \cdot
    \vartheta 
     \begin{bmatrix}
         \vec{j}_2 \\ 0
     \end{bmatrix}(\vec{z}_2, {\bf{N}}_2 \Omega)
= \sum_{\vec{l}_1, \vec{l}_2 \in \mathbb{Z}^2} e^{\pi i (\vec{\bm{j}} +\vec{\bm{l}})^T \cdot \bm{Q} \cdot (\vec{\bm{j}} +\vec{\bm{l}})} e^{2 \pi i(\vec{\bm{j}} +\vec{\bm{l}})^T \cdot \vec{\bm{z}}}  
\end{align}
where we defined
\begin{equation}
\vec{\bm{j}} +\vec{\bm{l}}= 
 \begin{pmatrix}
 \vec{j}_1 + \vec{l}_1 \\
 \vec{j}_2 + \vec{l}_2
 \end{pmatrix},\quad 
\vec{\bm{z}} = 
  \begin{pmatrix}
  \vec{z}_1 \\ \vec{z}_2
  \end{pmatrix}, \quad 
\bm{Q} = 
  \begin{pmatrix}
  {\bf{N}}_1 \Omega & 0 \\
    0 & {\bf{N}}_2 \Omega
  \end{pmatrix}.
\end{equation}
We introduce a transformation matrix:
\begin{equation}
T = 
 \begin{pmatrix}
 1 & 1 \\
 \alpha {\bf{N}}_1^{-1} & - \alpha {\bf{N}}_2^{-1}
 \end{pmatrix}, \quad
 T^{-1} = ({\bf{N}}_1^{-1} + {\bf{N}}_2^{-1})^{-1}
  \begin{pmatrix}
 {\bf{N}}_2^{-1} & \alpha^{-1} \\
 {\bf{N}}_1^{-1} & - \alpha^{-1}
 \end{pmatrix},
\end{equation}
where $\alpha$ is a $2 \times 2$ regular matrix. Then we have
\begin{align}
\begin{aligned}
\bm{Q}' &:= T \bm{Q} T^T =
    \begin{pmatrix}
({\bf{N}}_1 + {\bf{N}}_2) \Omega & 0 \\ 
0 & \alpha [{\bf{N}}_1^{-1} ({\bf{N}}_1 + {\bf{N}}_2) {\bf{N}}_2^{-1} \Omega^T ] \alpha^T
    \end{pmatrix},  \\
T \vec{\bm{z}} &= 
  \begin{pmatrix}
  \vec{z}_1 + \vec{z}_2 \\
  \alpha {\bf{N}}_1^{-1} \vec{z}_1 - \alpha {\bf{N}}_2^{-1} \vec{z}_2
  \end{pmatrix}, \\
(\vec{\bm{j}} +\vec{\bm{l}})^T T^{-1} &= 
 \begin{pmatrix}
 [(\vec{j}_1 + \vec{l}_1) {\bf{N}}_1 + (\vec{j}_2 + \vec{l}_2) {\bf{N}}_2] ({\bf{N}}_1 + {\bf{N}}_2)^{-1} \\
 [(\vec{j}_1 + \vec{l}_1) - (\vec{j}_2 + \vec{l}_2)] {\bf{N}}_1 ({\bf{N}}_1 + {\bf{N}}_2)^{-1} {\bf{N}}_2 \alpha^{-1}
 \end{pmatrix}.
 \end{aligned}
\end{align}
Eq.(\ref{eq: theta_product}) can be written as 
\begin{align}
\begin{aligned}
\label{eq: LHS}
     &\vartheta 
     \begin{bmatrix}
         \vec{j}_1 \\ 0
     \end{bmatrix}(\vec{z}_1, {\bf{N}}_1 \Omega) 
 \cdot
    \vartheta 
     \begin{bmatrix}
         \vec{j}_2 \\ 0
     \end{bmatrix}(\vec{z}_2, {\bf{N}}_2 \Omega)
= \sum_{\vec{l}_1, \vec{l}_2 \in \mathbb{Z}^2} e^{\pi i (\vec{\bm{j}} +\vec{\bm{l}})^T T^{-1} \cdot \bm{Q}' \cdot (T^{-1})^T(\vec{\bm{j}} +\vec{\bm{l}})} e^{2 \pi i(\vec{\bm{j}} +\vec{\bm{l}})^T T^{-1} \cdot T \vec{\bm{z}}} \\
& = \sum_{\vec{l}_1, \vec{l}_2 \in \mathbb{Z}^2} e^{\pi i  [(\vec{j}_1 + \vec{l}_1) {\bf{N}_1} + (\vec{j}_2 + \vec{l}_2) {\bf{N}}_2] ({\bf{N}}_1 + {\bf{N}}_2)^{-1} \cdot ({\bf{N}}_1 + {\bf{N}}_2) \Omega \cdot {({\bf{N}}_1 + {\bf{N}}_2)^{-1}}^T [{{\bf{N}}_1}^T (\vec{j}_1 + \vec{l}_1) + {{\bf{N}}_2}^T (\vec{j}_2 + \vec{l}_2) ]} \\
& \cdot e^{2 \pi i [(\vec{j}_1 + \vec{l}_1) {\bf{N}}_1 + (\vec{j}_2 + \vec{l}_2) {\bf{N}}_2] ({\bf{N}}_1 + {\bf{N}}_2)^{-1}\cdot (\vec{z}_1 + \vec{z}_2)} \\
& \cdot e^{\pi i  [(\vec{j}_1 + \vec{l}_1) - (\vec{j}_2 + \vec{l}_2)] {\bf{N}}_1 ({\bf{N}}_1 + {\bf{N}}_2)^{-1} {\bf{N}}_2 \alpha^{-1} \cdot \alpha [ {\bf{N}}_1^{-1} ({\bf{N}}_1 + {\bf{N}}_2) {\bf{N}}_2^{-1} \Omega^T ] \alpha^T \cdot {\alpha^{-1}}^{T} {\bf{N}}_2^T {({\bf{N}}_1 + {\bf{N}}_2)^{-1}}^T {\bf{N}}_1^T  [(\vec{j}_1 + \vec{l}_1) - (\vec{j}_2 + \vec{l}_2)]} \\
& \cdot e^{2 \pi i [(\vec{j}_1 + \vec{l}_1) - (\vec{j}_2 + \vec{l}_2)] {\bf{N}}_1 ({\bf{N}}_1 + {\bf{N}}_2)^{-1} {\bf{N}}_2 \alpha^{-1} \cdot \alpha [ {\bf{N}}_1^{-1} \vec{z}_1 - {\bf{N}}_2^{-1} \vec{z}_2] }.
\end{aligned}
\end{align}
We show that Eq.(\ref{eq: LHS}) is equal to 
\begin{align}
\begin{aligned}
\label{eq: RHS}
    & \sum_{\vec{l}_3, \vec{l}_4 \in \mathbb{Z}^2} \sum_{\vec{m}} e^{\pi i [(\vec{j}_1 {\bf{N}}_1 + \vec{j}_2 {\bf{N}}_2 + \vec{m} {\bf{N}}_1) ({\bf{N}}_1 + {\bf{N}}_2)^{-1} + \vec{l}_3] \cdot ({\bf{N}}_1 + {\bf{N}}_2) \Omega \cdot [({{\bf{N}}_1 + {\bf{N}}_2)^{-1}}^T ({\bf{N}}_1^T \vec{j}_1 + {\bf{N}}_2^T \vec{j}_2 + {\bf{N}}_1^T \vec{m}) + \vec{l}_3] } \\
   & \cdot e^{2 \pi i [(\vec{j}_1 {\bf{N}}_1 + \vec{j}_2 {\bf{N}}_2 + \vec{m} {\bf{N}}_1) ({\bf{N}}_1 + {\bf{N}}_2)^{-1} + \vec{l}_3] \cdot (\vec{z}_1 + \vec{z}_2)}  \\
   & \cdot e^{[(\vec{j}_1 - \vec{j}_2 + \vec{m}) \frac{{\bf{N}}_1 ({\bf{N}}_1 + {\bf{N}}_2)^{-1} {\bf{N}}_2}{{\rm lcm}(D_1, D_2)} P^{-1} + \vec{l}_4] \cdot ({\rm lcm}(D_1, D_2))^2 P {\bf{N}}_1^{-1} ({\bf{N}}_1 + {\bf{N}}_2) {\bf{N}}_2^{-1} \Omega^T P^T \cdot [{P^{-1}}^T \frac{{\bf{N}}_2^T {({\bf{N}}_1 + {\bf{N}}_2)^{-1}}^T {\bf{N}}_1^T}{{\rm lcm}(D_1, D_2)} (\vec{j}_1 - \vec{j}_2 + \vec{m}) + \vec{l}_4]} \\
   & \cdot e^{2 \pi i [(\vec{j}_1 - \vec{j}_2 + \vec{m}) \frac{{\bf{N}}_1 ({\bf{N}}_1 + {\bf{N}}_2)^{-1} {\bf{N}}_2}{{\rm lcm}(D_1, D_2)} P^{-1} + \vec{l}_4] \cdot {\rm lcm}(D_1, D_2) P ({\bf{N}}_1^{-1} \vec{z}_1 - {\bf{N}}_2^{-1} \vec{z}_2)},
\end{aligned}
\end{align}
if we take ${\alpha} = {\rm lcm}(D_1, D_2) P,\ P \in GL(2, \mathbb{Z})$. We verify the equality of Eqs.(\ref{eq: LHS}) and (\ref{eq: RHS}) in three steps. 

\paragraph{${\rm (I)}\ \vec{l}_1 = \vec{l}_2:$}\ \\
Firstly, terms in Eq.(\ref{eq: LHS}) with $\vec{l}_1 = \vec{l}_2$ correspond to those in Eq.(\ref{eq: RHS}) specified by the following relations, 
\begin{equation}
    \vec{l}_3 = \vec{l}_2,\quad   \vec{l}_4 = 0,\quad  \vec{m} = 0.
\end{equation}

\paragraph{${\rm (II)}\ \vec{l}_1 = \vec{l}_2 + \vec{l}_4 \alpha {\bf{N}}_2^{-1} ({\bf{N}}_1 + {\bf{N}}_2) {\bf{N}}_1^{-1},\ (\vec{l}_4 \neq 0):$}\ \\
Secondly, we focus on the case when the difference of $\vec{l}_1$ and $\vec{l}_2$ is given by
\begin{equation}
\label{eq: l_4}
    \vec{l}_1 - \vec{l}_2 = \vec{l}_4 \alpha {\bf{N}}_2^{-1} ({\bf{N}}_1 + {\bf{N}}_2) {\bf{N}}_1^{-1},\quad (\vec{l}_4 \in \mathbb{Z}^2/\{0\}).
\end{equation}
We can write Eq.(\ref{eq: l_4}) as
\begin{equation}
    \vec{l}_1 = \vec{l}_2 + \vec{l}_4 \alpha  ({\bf{N}}_1^{-1} + {\bf{N}}_2^{-1}).
\end{equation}
Note that $\vec{l}_i, (i=1,2,3,4) \in \mathbb{Z}^2$, and ${\bf{N}}_j, (j=1,2)$ are integer matrices. We can see that the following choice of $\alpha$ is possible, 
\begin{equation}
    \alpha = {\rm lcm}(D_1, D_2) P, \quad P \in GL(2, \mathbb{Z}).
\end{equation}
Then the correspondence of terms between Eqs.(\ref{eq: LHS}) and (\ref{eq: RHS}) is given by
\begin{align}
\label{eq: difference}
    \vec{l}_1 - \vec{l}_2 &= \vec{l}_4 {\rm lcm}(D_1, D_2) P {\bf{N}}_2^{-1} ({\bf{N}}_1 + {\bf{N}}_2) {\bf{N}}_1^{-1}, \\
    \label{eq: l_2}
    \vec{l}_2 &= - \vec{l}_4 {\rm lcm}(D_1, D_2) P {\bf{N}}_2^{-1} + \vec{l}_3, \\
\vec{m} &= 0.
\end{align}
For an arbitrary choice of $\vec{l}_3 \in \mathbb{Z}^2$ and  $\vec{l}_4 \in \mathbb{Z}^2 / \{0 \}$, we obtain unique $\vec{l}_2 \in \mathbb{Z}^2$ from Eq.(\ref{eq: l_2}) and the difference $\vec{l}_1 - \vec{l}_2 (\neq 0)$ from Eq.(\ref{eq: difference}). 

\paragraph{${\rm (III)}\ \vec{l}_1 = \vec{l}_2 + \vec{l} \alpha {\bf{N}}_2^{-1} ({\bf{N}}_1 + {\bf{N}}_2) {\bf{N}}_1^{-1} + \vec{m}',\  (\vec{m}' \neq 0):$}\ \\
Lastly, we consider the case when the difference of $\vec{l}_1$ and $\vec{l}_2$  cannot be given by Eq.(\ref{eq: l_4}). We write the difference as
\begin{equation}
     \vec{l}_1 = \vec{l}_2 + \vec{l} \cdot  {\rm lcm}(D_1, D_2) P {\bf{N}}_2^{-1} ({\bf{N}}_1 + {\bf{N}}_2) {\bf{N}}_1^{-1}+ \vec{m}',
\end{equation} 
where $\vec{l} \in \mathbb{Z}^2$ and $\vec{m}'\in \mathbb{Z}^2 / \{ 0\}$.
Then the correspondence with Eq.(\ref{eq: RHS}) is 
\begin{equation}
    \vec{l}_3 = \vec{l}_2,\quad \vec{l}_4 = \vec{l}, \quad \vec{m} = \vec{m}'.
\end{equation}
The sum over $\vec{m}$ is finite. To understand it, suppose $\vec{m}'$ is large enough such that we have $\vec{L} \in \mathbb{Z}^2 / \{ 0 \}$ satisfying
\begin{equation}
     \vec{m}' = \vec{L} \cdot {\rm lcm}(D_1, D_2) P {\bf{N}}_1^{-1}({\bf{N}}_1 + {\bf{N}}_2){\bf{N}}_2^{-1} = \vec{L} \cdot {\rm lcm}(D_1, D_2) P ({\bf{N}}_1^{-1} + {\bf{N}}_2^{-1}).
\end{equation}
Then we notice that this is what we have already considered in (II). Thus, we can understand Eq.(\ref{eq: cell_m}). Let us summarize what we have looked at. (I) and (III) show that we have corresponding $\vec{l}_3, \vec{l}_4 \in \mathbb{Z}^2$, and $\vec{m}$ for $\forall \vec{l}_1, \vec{l}_2 \in \mathbb{Z}^2$. Conversely, (II) and (III) show that we have corresponding $\vec{l}_1, \vec{l}_2 \in \mathbb{Z}^2$ for $\forall \vec{l}_3, \vec{l}_4 \in \mathbb{Z}^2$, and $\forall \vec{m}$. Consequently, we obtain Eq.(\ref{eq: theta_id}).
$\Box$

\subsection{Computation of Yukawa couplings}
We evaluate the overlap integral of the wave functions including Scherk-Schwarz phases as shown in Eq.(\ref{eq: Yukawa_general}).
First, we note the following property of the Riemann theta-function:
\begin{equation}
\vartheta
 \begin{bmatrix}
 \vec{j} + \vec{\alpha}_L {\bf N}^{-1}_L \\
 - \vec{\beta}_L
 \end{bmatrix}({\bf{N}}_L \vec{z}, {\bf{N}}_L\Omega)
 = 
 \vartheta
 \begin{bmatrix}
 \vec{j} + \vec{\alpha}_L {\bf N}^{-1}_L \\
 0
 \end{bmatrix}({\bf{N}}_L \vec{z} - \vec{\beta}_L, {\bf{N}}_L\Omega).
\end{equation}
By use of Eq.(\ref{eq: theta_id}), we get
\begin{align}
\begin{aligned}
\label{eq: theta_id_SS}
    &\vartheta
 \begin{bmatrix}
 \vec{i} + \vec{\alpha}_L {\bf N}^{-1}_L \\
 - \vec{\beta}_L
 \end{bmatrix}({\bf{N}}_L \vec{z}, {\bf{N}}_L\Omega) \cdot 
 \vartheta
 \begin{bmatrix}
 \vec{j} + \vec{\alpha}_R {\bf N}^{-1}_R \\
 - \vec{\beta}_R 
 \end{bmatrix}({\bf{N}}_R \vec{z}, {\bf{N}}_R\Omega) \\
 &= \sum_{\vec{m}} 
    \vartheta 
     \begin{bmatrix}
     ((\vec{i} + \vec{\alpha}_L {\bf N}^{-1}_L ){\bf{N}}_L + (\vec{j} + \vec{\alpha}_R {\bf N}^{-1}_R) {\bf{N}}_R + \vec{m} {\bf{N}}_L) ({\bf{N}}_L + {\bf{N}}_R)^{-1} \\
     0 
     \end{bmatrix} \\
     &(({\bf{N}}_L + {\bf{N}}_R )\vec{z} - \vec{\beta}_L - \vec{\beta}_R, ({\bf{N}}_L+{\bf{N}}_R)\Omega) \\
& \times  
     \vartheta 
     \begin{bmatrix}
     (\vec{i} + \vec{\alpha}_L{\bf{N}}_L^{-1} - \vec{j}  - \vec{\alpha}_R {\bf{N}}_R^{-1} + \vec{m}  ) 
     \frac{{\bf{N}}_L ({\bf{N}}_L + {\bf{N}}_R)^{-1} {\bf{N}}_R}{{\rm lcm}(D_L, D_R)}P^{-1} \\
     0 
     \end{bmatrix} \\
&
({\rm lcm}(D_L, D_R) P (-{\bf{N}}_L^{-1} \vec{\beta}_L + {\bf{N}}_R^{-1} \vec{\beta}_R) , ({\rm lcm}(D_L, D_R))^2 P {\bf{N}}_L^{-1} ({\bf{N}}_L + {\bf{N}}_R) {\bf{N}}_R^{-1} \Omega^T P^T).
\end{aligned}
\end{align}
Then we can express the product of two zero-mode wave functions of left- and right-handed fermions as 
\begin{align}
\begin{aligned}
 &\psi^{(\vec{i}+\vec{\alpha}_L {\bf{N}}_L^{-1}, \vec{\beta}_L)} (\vec{z}, \Omega) \cdot \psi^{(\vec{j}+\vec{\alpha}_R {\bf{N}}_R^{-1}, \vec{\beta}_R)} (\vec{z}, \Omega) \\ &= \frac{\mathcal{N}_L \mathcal{N}_R}{\mathcal{N}_H} \sum_{\vec{m}} \psi^{(\{(\vec{i} + \vec{\alpha}_L {\bf N}^{-1}_L ){\bf{N}}_L + (\vec{j} + \vec{\alpha}_R {\bf N}^{-1}_R) {\bf{N}}_R + \vec{m} {\bf{N}}_L\} {\bf{N}}_H^{-1} , 0) } (\vec{z} - (\vec{\beta}_L + \vec{\beta}_R){\bf{N}}_H ^{-1},  \Omega)  \\
& \times  
     \vartheta 
     \begin{bmatrix}
     (\vec{i} + \vec{\alpha}_L{\bf{N}}_L^{-1} - \vec{j}  - \vec{\alpha}_R {\bf{N}}_R^{-1} + \vec{m} ) 
     \frac{{\bf{N}}_L {\bf{N}}_H^{-1} {\bf{N}}_R}{{\rm lcm}(D_L, D_R)}P^{-1} \\
     0 
     \end{bmatrix} \\
&
({\rm lcm}(D_L, D_R) P (-{\bf{N}}_L^{-1} \vec{\beta}_L + {\bf{N}}_R^{-1} \vec{\beta}_R) , {\rm lcm}(D_L, D_R)^2 P {\bf{N}}_L^{-1} {\bf{N}}_H {\bf{N}}_R^{-1} \Omega^T P^T),
\end{aligned}
\end{align}
where $\mathcal{N}_A,\ (A=L,H,R)$ are normalization constants of wave functions in the three sectors. Here we used the relation ${\bf{N}}_L + {\bf{N}}_R = {\bf{N}}_H$. Now, we can easily perform the overlap integral in Eq.(\ref{eq: overlap}) by use of the orthonormality condition
\begin{equation}
\label{eq: orthogonal}
    \int_{T^4} d^2z d^2{\bar{z}}\  \psi^{(\vec{j}+\vec{\alpha} {\bf{N}}^{-1}, \vec{\beta})} 
    (\vec{z}, \Omega)
    \left(\psi^{(\vec{k}+\vec{\alpha} {\bf{N}}^{-1}, \vec{\beta})}
     (\vec{z}, \Omega) \right)^* = \delta_{\vec{j}, \vec{k} + \vec{l}}\ ,
\end{equation}
where $\vec{l} \in \mathbb{Z}^2$ \cite{Cremades:2004wa}.
By applying Eq.(\ref{eq: orthogonal}), we have
\begin{align}
\begin{aligned}
    &\int_{T^4} d^2z d^2 \bar{z}\ 
   \psi^{(\{(\vec{i} + \vec{\alpha}_L {\bf N}^{-1}_L ){\bf{N}}_L + (\vec{j} + \vec{\alpha}_R {\bf N}^{-1}_R) {\bf{N}}_R + \vec{m} {\bf{N}}_L\} {\bf{N}}_H^{-1} , 0) } (\vec{z} - (\vec{\beta}_L + \vec{\beta}_R){\bf{N}}_H ^{-1},  \Omega) 
 \cdot \left( \psi^{(\vec{k}+\vec{\alpha}_H {\bf{N}}_H^{-1}, \vec{\beta}_H)} (\vec{z}, \Omega) \right)^* \\
    &=\int_{T^4} d^2z d^2 \bar{z}\ \psi^{(\{(\vec{i} + \vec{\alpha}_L {\bf N}^{-1}_L ){\bf{N}}_L + (\vec{j} + \vec{\alpha}_R {\bf N}^{-1}_R) {\bf{N}}_R + \vec{m} {\bf{N}}_L\} {\bf{N}}_H^{-1}, \vec{\beta}_H) } (\vec{z} ,  \Omega) \cdot \left( \psi^{(\vec{k}+\vec{\alpha}_H {\bf{N}}_H^{-1}, \vec{\beta}_H)} (\vec{z}, \Omega) \right)^* \\
    &= \delta_{((\vec{i} + \vec{\alpha}_L {\bf N}^{-1}_L ){\bf{N}}_L + (\vec{j} + \vec{\alpha}_R {\bf N}^{-1}_R) {\bf{N}}_R + \vec{m} {\bf{N}}_L ) {\bf{N}}_H^{-1} ,\vec{k}+\vec{\alpha}_H {\bf{N}}_H^{-1} + \vec{l}}\ ,
\end{aligned}
\end{align}
where $\vec{\beta}_H = \vec{\beta}_L + \vec{\beta}_R$ and $\vec{l} \in \mathbb{Z}^2$. From the above results, we immediately obtain Eqs.(\ref{eq: Yukawa_general}) and (\ref{eq: m_sum}).

\section{Details of Section 6}
\subsection{More about $\vec{m} \in \mathbb{Z}^2$ and $P \in GL(2, \mathbb{Z})$}
\label{m_and_P}
Here, we give a proof showing the choice of $\vec{m}$ is independent of $P \in GL(2,\mathbb{Z})$. 
Suppose we obtained $M=\{\vec{m}_1, \vec{m}_2, \cdots ,\vec{m}_{{\rm det}{\bf N}_H} \}$ from the definition of $\vec{m}$ in Eq.(\ref{eq: m_def}) when $P=I$. Similarly, suppose we obtained $M' = \{\vec{m}'_1, \vec{m}'_2, \cdots ,\vec{m}'_{{\rm det}{\bf N}_H} \}$ when $P=P_1$. There is a one to one correspondence 
\begin{equation}
\label{eq: m_mod}
    \vec{m}_a' \equiv \vec{m}_b,\quad \pmod{\vec{e}_n ({\rm det}{\bf N}_H){\bf N}_H^{-1}},
\end{equation}
between each element of $M$ and $M'$. 

Firstly, let us show that for any $\vec{m}'_a \in M$, there exists a unique $\vec{m}_b \in M$ such that Eq.(\ref{eq: m_mod}) is satisfied.
It us obvious that such $\vec{m}_b$ exists at least one. To show the uniqueness, assume we had $\vec{m}_a' \equiv \vec{m}_b$ and $\vec{m}_a' \equiv \vec{m}_c, \pmod{\vec{e}_n ({\rm det}{\bf N}_H){\bf N}_H^{-1}}$ where $b \neq c,\ (1 \leq b < c \leq {\rm det}{\bf N}_H)$. However, we find a contradiction since $\vec{m}_b = \vec{m}_c$ must be true because none of the elements of $M$ are identified modulo $\vec{e}_n ({\rm det}{\bf N}_H){\bf N}_H^{-1}$.

Secondly, let us show that for any $\vec{m}_b \in M$, there exists a unique $\vec{m}'_a \in M'$ such that Eq.(\ref{eq: m_mod}) is satisfied. Let us
assume we had $\vec{m}_d \equiv \vec{m}'_e$ and $\vec{m}_d \equiv \vec{m}'_f,\ \pmod{\vec{e}_n ({\rm det}{\bf N}_H){\bf N}_H^{-1}}$ where $e \neq f, (1 \leq e < f \leq {\rm det}{\bf N}_H )$. We should notice that the lattice points generated by $\vec{e}_n P ({\rm det}{\bf N}_H) {\bf N}_H^{-1}$ are independent of $P$. This means none of the elements of $M'$ are identified with each other modulo ${\vec{e}_n ({\rm det}{\bf N}_H){\bf N}_H^{-1}}$. 
Then, $\vec{m}'_e = \vec{m}'_f$ must be true, but this is a contradiction. Thus, we showed the claim of Eq.(\ref{eq: m_mod}).

Next, we verify that our expression of the Yukawa coupling constants in Eq.(\ref{eq: Yukawa_Types}) is invariant under the replacement, $\vec{m} \rightarrow \vec{m} + \vec{e}_n  ({\rm det}{\bf N}_H) {\bf N}_H^{-1}$. The characteristic only receives a change by an integer vector
\begin{equation}
    \vec{e}_n  P^{-1} \in \mathbb{Z}^2,
\end{equation}
which is trivial for the Riemann theta-function. The selection rule is also unaffected. This is because the replacement only induces a change which can be absorbed by the term $\vec{l} {\bf N}_H$. Eq.(\ref{eq: useful}) may be useful to show this. 
Thus, we can choose $\vec{m}$ independent of $P$ because Yukawa coupling constants are invariant.

\subsection{Alignment of $\vec{m}$}
\label{m_and_Type}
Here, we explain why knowing the Type of ${\bf N}_H$ can determine the alignment of $\vec{m}$ in Eq.(\ref{eq: selection}).
We consider ${\bf N}_H$ parameterized as
\begin{equation}
    {\bf N}_H =
    \begin{pmatrix}
    n_{11} & n_{12} \\
    n_{21} & n_{22}
    \end{pmatrix}.
\end{equation}
Then we obtain,
\begin{equation}
    ({\rm det}{\bf N}_H) {\bf N}_H^{-1} =
    \begin{pmatrix}
    n_{22} & - n_{12} \\
    - n_{21} & n_{11}
    \end{pmatrix}.
\end{equation}
$\vec{m}$ were defined as $\mathbb{Z}^2$ vectors inside the cell spanned by $\vec{e}_n  ({\rm det}{\bf N}_H){\bf N}_H^{-1}$. 

Suppose we had
${\rm gcd}(n_{21},n_{22})=1$. 
Then we can relocate $\vec{m}$ as 
\begin{equation}
\label{eq: line}
    (0,0), (0,1), \cdots , (0,({\rm det}{\bf N})-1),
\end{equation}
if we allow the shift by $\vec{e}_n  ({\rm det}{\bf N}_H){\bf N}_H^{-1}$. Note that under such shifts of $\vec{m}$ the Yukawa couplings in Eq.(\ref{eq: Yukawa_Types}) were invariant as we explained in Appendix \ref{m_and_P}.
Eq.(\ref{eq: line}) is explained as follows. By taking a linear combination of $\vec{e}_n ({\rm det}{\bf N}_H){\bf N}_H^{-1}$, we have a vector
\begin{equation}
    \vec{v} = 
    \begin{pmatrix}
    v_1 \\ v_2
    \end{pmatrix}=
    \begin{pmatrix}
    s n_{22} - t n_{21} \\
    -s n_{12} + t n_{11}
    \end{pmatrix},\quad s, t \in \mathbb{Z}.
\end{equation}
Let us choose $s$ and $t$ to satisfy $v_1=sn_{22} - t n_{21}=0$.
Since $n_{11}$ and $n_{21}$ are coprime, general solutions are 
\begin{equation}
    s =  n_{21}l,\ t= n_{22} l,\quad l\in \mathbb{Z}. 
\end{equation}
If we take $l=1$,
\begin{equation}
\label{eq: v_shift}
    \vec{v} = \begin{pmatrix}
    0 \\
    {\rm det}{\bf N}_H
    \end{pmatrix},
\end{equation}
which is the smallest unit of identification along $\vec{e}_2 = (0,1)$. All of the integer points on Eq.(\ref{eq: v_shift}) are not identified with each other. Thus, we verified the claim of Eq.(\ref{eq: line}).

In the above argument, we used the condition ${\rm gcd}(n_{21},n_{22})=1$.  As we showed in Appendix \ref{gcd_proof}, intersection matrices in the same Type have the same ${\rm gcd}(n_{11},n_{12})$ and ${\rm gcd}(n_{21},n_{22})$. Thus, once we identified the Type of ${\bf N}_H$, we have information to determine possible alignments of $\vec{m}$ modulo $\vec{e}_n  ({\rm det}{\bf N}_H){\bf N}_H^{-1}$. Although we have only considered the case when ${\rm gcd}(n_{21},n_{22})=1$, we can extend the similar analysis for general cases.

\subsection{Greatest common divisor and Type}
\label{gcd_proof}
Here, we prove intersection matrices $\bf N$ in the same Type have the same ${\rm gcd}(n_{11},n_{12})$ and ${\rm gcd}(n_{21},n_{22})$, where
\begin{equation}
\label{eq: N_gcd}
    {\bf N} = 
    \begin{pmatrix}
    n_{11} & n_{12} \\
    n_{21} & n_{22}
    \end{pmatrix}.
\end{equation}
It suffices to confirm actions of  ${SL(2,\mathbb{Z})}$ generators in Eq.(\ref{eq: SL2Z_generators}). 

First, consider the S transformation,
\begin{equation}
    {\bf N} S =
    \begin{pmatrix}
    n_{12} & - n_{11} \\
    n_{22} & - n_{21}
    \end{pmatrix}.
\end{equation}
We find that the greatest common divisors of both rows are preserved. 

Second, consider the T transformation,
\begin{equation}
    {\bf N} T =
    \begin{pmatrix}
    n_{11} & n_{11}+n_{12} \\
    n_{21} & n_{21}+n_{22}
    \end{pmatrix}.
\end{equation}
It is obvious that $g' := {\rm gcd}(n_{11},n_{11}+n_{12}) \geq {\rm gcd}(n_{11}, n_{12})$ holds. To show that it must be an equality, suppose $g' := {\rm gcd}(n_{11},n_{11}+n_{12}) > {\rm gcd}(n_{11}, n_{12})$. Then we have
\begin{equation}
    \frac{n_{11}+n_{12}}{g'} = \frac{n_{11}}{g'} + \frac{n_{12}}{g'} ,
\end{equation}
where $\frac{n_{11}+n_{12}}{g'} \in {\mathbb{Z}}$ and $\frac{n_{11}}{g'} \in \mathbb{Z}$. Thus, $\frac{n_{12}}{g'} \in \mathbb{Z}$ must be true and $n_{11}$ and ${n_{12}}$ have a common factor $g'$. However, this is a contradiction. Therefore, we proved 
\begin{equation}
    g'= {\rm gcd}(n_{11},n_{11}+n_{12}) = {\rm gcd}(n_{11}, n_{12}).
\end{equation}
Similarly, we can verify ${\rm gcd}(n_{21},n_{21}+n_{22}) = {\rm gcd}(n_{21}, n_{22})$. 

It is also possible to show that intersection matrices $\bf N$ in the same L-Type have the same ${\rm gcd}(n_{11},n_{21})$ and ${\rm gcd}(n_{12},n_{22})$. 

%%%%%%%%%%%%%%%%%%%%%%%%%%%%%%%%%%%%%%%%%

 \end{document}